%% file: manuscript.tex
\documentclass[12pt,a4paper]{article} 
\usepackage[utf8]{inputenc}
\usepackage[T1]{fontenc}
\usepackage[ngerman,english]{babel}
\usepackage{lmodern}
\usepackage[a4paper,inner=2.5cm,bottom=2.5cm,outer=2.5cm,top=2.5cm]{geometry}
\usepackage{setspace} % vertical spacing
\usepackage[]{graphicx}
\usepackage{epstopdf}       
\usepackage{csquotes}
\usepackage[style=authoryear,backend=bibtex8,sorting=nyt,doi=false,isbn=false,url=true,abbreviate=true,eprint=false,natbib,
maxbibnames=10
]{biblatex}
\usepackage{xcolor}
\usepackage{rotating}
\usepackage{float}
\usepackage[section]{placeins}
\usepackage{tabularx,calc}
\usepackage{makecell}
\usepackage{amsmath,amsfonts,amssymb,amsthm} 
\usepackage{bm} % defined \bm command for bold maths symbols
\usepackage{bbm}
\usepackage{xspace} 
\usepackage{caption}
\usepackage[justification=centering]{subcaption}
\usepackage{multirow,array,varwidth}
\usepackage[hidelinks]{hyperref}
\usepackage{enumitem}
\usepackage{xr}
\usepackage{minitoc}
\usepackage{booktabs}
\usepackage{pifont}
\usepackage{threeparttable} % for tables with table notes
\usepackage[colorinlistoftodos]{todonotes}
\usepackage{siunitx}
\usepackage{array}
\usepackage{mathtools}
\newcolumntype{H}{>{\setbox0=\hbox\bgroup}c<{\egroup}@{}} % hide column
\usepackage{tcolorbox}
\usepackage{comment}
\usepackage{listings}
\usepackage{color}
\usepackage{scalerel} 

\usepackage{epigraph}

\usepackage{algorithm}
\usepackage{algorithmic}

\input{partialinput}

\AtEveryBibitem{\clearfield{note}}
\AtEveryBibitem{\clearfield{month}}
\AtEveryBibitem{\clearfield{urlyear}}
\AtEveryBibitem{\clearlist{language}}
\renewbibmacro{in:}{%
  \ifentrytype{article}{}{\printtext{\bibstring{in}\intitlepunct}}
}
\AtEveryBibitem{%
\ifentrytype{article}{\clearfield{url}}%
}

\input{notation}

\input{theory_envs} % boxes are defined here
    
\title{An Introduction to \\ Double/Debiased Machine Learning
}
\author{Achim Ahrens \and Victor Chernozhukov \and Christian Hansen \and Damian Kozbur \and Mark Schaffer \and Thomas Wiemann\thanks{We thank Thiago Cacicedo dos Santos, Francesca Codega, Sara Drango, Teresa Freitas Monteiro, Samuel Higbee, Štěpán Jurajda, Rafael Lalive, Hugo Lopez, Claudia Marangon, Moritz Marbach, Max Maydanchik, Eoin McLaughlin, and Alessandra Stampi-Bombelli for reading and providing helpful comments on drafts of the paper. We also thank seminar participants at the Luxembourg Institute of Socio-Economic Research (LISER) and participants at the \emph{Tools, Data, and Methods Workshop} at the University of Exeter. We further thank the editor, David Romer, and five anonymous referees for constructive comments and suggestions. Computational resources were provided by ETH Z\"urich's DeSciL, and the e-INFRA CZ project (ID:90254), supported by the Ministry of Education, Youth and Sports of the Czech Republic. We used OpenAI's GPT-4o, GPT-5, GPT-5.2 and Anthropic's Claude Opus 4.5 to assist with proofreading and editing during the preparation of this manuscript. All errors are our own.}}
\date{\today}

\begin{document}

\emergencystretch 3em

\maketitle

% Another attempt. Spelled out nuisance function. 147 words. 
This paper provides an introduction to Double/Debiased Machine Learning (DML). DML is a general approach to performing inference about a target parameter in the presence of nuisance functions: objects that are needed to identify the target parameter but are not of primary interest. Nuisance functions arise naturally in many settings, such as when controlling for confounding variables or leveraging instruments. The paper describes two biases that arise from nuisance function estimation and explains how DML alleviates these biases. Consequently, DML allows the use of flexible methods, including machine learning tools, for estimating nuisance functions, reducing the dependence on auxiliary functional form assumptions and enabling the use of complex non-tabular data, such as text or images. We illustrate the application of DML through simulations and empirical examples. We conclude with a discussion of recommended practices. A companion website includes additional examples with code and references to other resources.

%\newpage

\section{Introduction}

A large share of empirical research in economics aims to provide insights into the statistical relationships among two or more variables. For example, a common research goal is to understand the causal impact of a policy on economic outcomes.  Target parameters summarizing these relationships, including average treatment effects or regression coefficients, frequently depend on \emph{nuisances}---auxiliary objects that must be accounted for to identify the parameter of interest but are not themselves of primary interest, such as regression coefficients on control variables.

As a concrete example that we will revisit in our empirical illustrations, consider \citet{dube2020}, who study monopsony power on the online platform MTurk using partially linear regression:
\begin{align*}
Y &= \theta_0 D + g_0(X) + \varepsilon
\end{align*}
where $Y$ is the logarithm of the time it takes for a posted job to be filled, $D$ is the logarithm of the reward of the job, $X$ denotes observed features of tasks including their type and complexity, and $\varepsilon$ is assumed to be  uncorrelated with $D$ and mean independent of $X$, i.e., $\Ep[D \varepsilon] = 0$ and $\Ep[\varepsilon|X] = 0$. Their target parameter is the regression coefficient $\theta_0$, which they interpret as a measure of the negative labor supply elasticity. %\citet{dube2020} emphasize that heterogeneity in tasks, $X$, needs to be accounted for in order for $\theta_0$ to have a meaningful economic interpretation.

In this example, $g_0(\cdot)$ is a nuisance function. We are not primarily interested in how task features, outside of reward, relate to the outcome. However, \citet{dube2020} emphasize that task heterogeneity needs to be accounted for to meaningfully interpret $\theta_0$.

If $g_0(\cdot)$ were known, estimation of $\theta_0$ could proceed by regressing $Y - g_0(X)$ onto $D$, which is equivalent to estimation based on the moment condition
\begin{align*}
    \Ep[m(W;\theta_0,g_0)] = \Ep[(Y-g_0(X)-D\theta_0)D] = 0
\end{align*}
where $W = (Y,D,X)$ denote observed random variables. Of course, $g_0(\cdot)$ will typically not be known. A common strategy to simplify the problem is to \emph{assume} that $g_0(X) = X'\beta_0$ with unknown coefficient $\beta_0$, in which case the model reduces to the familiar multiple regression model. When the dimension of $X$ is much smaller than the sample size, estimation would then proceed by ordinary least squares (OLS) of $Y$ on $D$ and $X$. 

Even in the simple multiple regression model, we have a nuisance parameter, $\beta_0$. Conventional regression estimates it jointly with the target parameter $\theta_0$. Alternatively, one can partial $X$ out from both $D$ and $Y$ to obtain residuals that isolate the variation identifying $\theta_0$. By the Frisch-Waugh-Lovell Theorem, regressing these residuals on each other yields an estimate of $\theta_0$ that is numerically equivalent to that obtained from regressing $Y$ on $D$ and $X$. This notion of ``partialling out'' is related to a broader principle for handling nuisance parameters that we emphasize throughout the review.

In the actual \citet{dube2020} example, some of the task characteristics are captured as text data, which makes it difficult to justify \emph{ad hoc} parametric assumptions like those in the linear model. Instead, \citet{dube2020} allow $g_0(\cdot)$ to be a flexible, \emph{high-dimensional} function rather than committing to a low-dimensional functional form. %That is, the \citet{dube2020} scenario involves a \emph{high-dimensional} nuisance object. 
In contrast to the low-dimensional linear case, where $g_0(X)=X'\beta_0$ reduces to estimating a small set of coefficients, estimation of $g_0(\cdot)$ must then also accommodate rich nonlinear variation.% and thus cannot be easily approximated by a known, low-dimensional parametric form.

This regression setup illustrates a broader template common in empirical research, in which the target parameter $\theta_0$ is defined as the solution to a moment condition:
\begin{align}\label{eq: intro moment}
    \theta_0: \: \Ep[m(W;\theta_0,\eta_0)] = 0.
\end{align}
Here, $m$ is a score (or moment) function and $W$ again denotes observed random variables. The parameter $\eta_0$ denotes a nuisance object, which is not of direct interest but is used to define $\theta_0$. $\eta_0$ is often high-dimensional. For example, $\eta_0$ will represent a vector of conditional expectation functions in many interesting cases. This \emph{semi-parametric} structure encompasses the regression example above, where the target parameter is the coefficient on $D$. It more generally applies to the estimation of many other canonical parameters, including average treatment effects, parameters in linear instrumental variables models, local average treatment effects, dynamic treatment effects in staggered adoption designs, and parameters in nonlinear structural models. 

High-dimensional nuisance parameters can arise in several ways: (i)~when the nuisance function depends on only a few covariates, controls, or instruments, but no parametric model is specified; (ii)~when there are many such variables, even under parametric assumptions; or (iii)~when numerous variables enter through unknown functions. High-dimensionality is increasingly common in applications using text or image data \citep[]{gentzkow2019b}, but it can also arise in simpler settings. Even a single continuous covariate may create a high-dimensional problem---for example, when identification relies on rainfall instruments that can be nonlinearly related to an endogenous regressor \citep[e.g.,][]{hidalgo2010a,gilchrist2016something,dustmann2017}.

As rich data become more common in applied research, there is growing appreciation that traditional functional form assumptions are often difficult to justify, motivating the use of more flexible tools for estimating nuisance parameters. As a result, there is increasing interest in using machine learning (ML) methods, which provide flexible tools for estimating high-dimensional nuisance parameters. A natural use of ML would then be to obtain nuisance parameter estimates, $\hat\eta$, and use these in place of $\eta_0$ in \eqref{eq: intro moment}. Specifically, we can define an estimator $\hat\theta$ of $\theta_0$ as a solution to the sample analog of \eqref{eq: intro moment}:
$$
\hat\theta:\: \frac{1}{n} \sum_{i=1}^{n} m(W_i;\hat\theta,\hat\eta) = 0
$$
where $W_i$ denote observed variables for observations $i = 1,...,n$. However, the resulting ``plug-in'' estimator $\hat\theta$---so-called because it ``plugs'' $\hat\eta$ in for $\eta_0$---can behave poorly and lead to misleading conclusions due to errors in estimating $\eta_0$ propagating into $\hat\theta$.%However, the resulting ``plug-in'' estimator of $\theta_0$ --- so-called because it ``plugs'' $\hat\eta$ in for $\eta_0$---can behave poorly and lead to misleading conclusions. The reason is that errors in estimating $\eta_0$ generally propagate into the estimator of $\theta_0$.

In settings with high-dimensional nuisance parameters, standard asymptotic approximations may fail due to two distinct forms of sensitivity to nuisance parameter estimation, termed \emph{regularization bias} and \emph{overfitting bias}. Both terms describe channels through which using an estimated nuisance parameter, $\hat{\eta}$, instead of the true but unknown $\eta_0$, can distort the behavior of the plug-in estimator $\hat\theta$. A leading manifestation of these distortions is bias, as the terminology suggests, but they more broadly invalidate conventional inference methods that fail to account for nuisance estimation. %Roughly speaking, regularization bias refers to the direct impact of estimation error in $\hat\eta$ on the plug-in estimator that results from the difference between $m(W;\theta_0,\hat\eta)$ and $m(W;\theta_0,\eta_0)$. %Overfitting bias refers to a more subtle issue introduced by statistical dependence between $\hat\eta$, which is a function of the data used in its estimation, and the data used to estimate $\theta_0$. 

Roughly speaking, regularization bias refers to the direct impact of estimation error in $\hat\eta$ on the plug-in estimator $\hat\theta$ that results from the difference between $m(W;\theta_0,\hat\eta)$ and $m(W;\theta_0,\eta_0)$. Overfitting bias refers to a more subtle issue. Because $\hat\eta$ is an estimator, it is itself a random function of the data. $\hat\eta$ is thus generally correlated with the observations $\{W_i\}_{i=1}^{n}$ also used in the estimating equation $\frac{1}{n}\sum_{i=1}^{n} m(W_i;\theta,\hat\eta)$. %, with a leading case being when the same data are used both to estimate $\hat\eta$ and evaluate the moment condition. 
When this dependence is strong, for example due to ``overfitting'', it %between $\hat\eta$ and $\{W_i\}_{i=1}^{n}$ 
may generate large differences between $\frac{1}{n}\sum_{i=1}^{n} m(W_i;\theta,\hat\eta)$ and $\frac{1}{n}\sum_{i=1}^{n} m(W_i;\theta,\eta_0)$, which results in poor performance of $\hat\theta$.

Both regularization and overfitting biases are major concerns in high-dimensional contexts. Accurately estimating high-dimensional $\eta_0$ is inherently difficult and often results in non-negligible errors. Further, estimation in high-dimensional settings typically relies on highly data-adaptive procedures---such as modern ML methods---which amplify the risk of overfitting bias. As such, mitigating these biases is the focus of a large and rapidly growing literature in statistics and econometrics that builds from classic ideas in semiparametric and nonparametric estimation. One method that provides a solution in a wide variety of empirical settings, and that is the topic of this review, is double/debiased machine learning \citep[henceforth DML;][]{chernozhukov2018}.

DML provides a blueprint for alleviating both regularization and overfitting bias. At its core, DML combines two classical ideas from the rich literature on semiparametric inference---using \emph{Neyman orthogonal scores}\footnote{Such scores are also referred to as orthogonal scores, orthogonal moments, locally robust moments, debiased moments, influence functions, and pathwise derivatives. We follow \citet{chernozhukov2018} and use the term ``Neyman orthogonal scores'' in homage to Neyman's early contributions, e.g., \citet{NeymanCalpha,Neyman1979}.} to alleviate regularization bias and using \emph{cross-fitting} to alleviate overfitting bias---in a common methodological framework. Neyman orthogonality ensures that plugging in estimates that are close to, but not exactly equal to, $\eta_0$ does not lead to large changes in the moment condition \eqref{eq: intro moment}.\footnote{
%Not all scores that serve to identify a parameter of interest satisfy Neyman orthogonality. 
Neyman orthogonality is not guaranteed for all scores that serve to identify a parameter of interest. %A prominent counterexample is the inverse propensity weighted score commonly used for treatment effect estimation. 
We show how to construct a Neyman orthogonal score from a given score in Appendix \ref{app:no_construction}.} 
Cross-fitting, a form of sample splitting, alleviates potential dependence between nuisance estimates $\hat{\eta}$ and parts of the data used for estimating the target parameter. Used in conjunction, DML's two core components significantly reduce the impact of nuisance estimation on estimates of the target parameter. However, high-quality estimation of the target parameter $\theta_0$ still requires nuisance parameters to be estimated sufficiently well. Consequently, theoretical results for DML assume specific convergence rate conditions on nuisance estimators. Many estimators, including ML methods, can satisfy these conditions. 

From a practical perspective, DML enables researchers to leverage a wide range of ML tools, making it particularly valuable in complex data scenarios involving numerous variables, images, or text data. Importantly, the benefits of ML also extend to traditional research settings with fewer covariates and conventional tabular data as they remove the need for researchers to commit beforehand to specific parametric (often linear) models. Furthermore, DML is easy to implement, applicable to a wide range of econometric settings, and readily available in existing software packages, including Stata, R, and Python \citep[e.g.,][]{DoubleML2021R,DoubleML2022Python,Ahrens2023_ddml,ddml_R}. DML thus has the potential to enhance the credibility of research findings in a broad spectrum of settings, either when used as a complementary robustness check or when the application necessitates the use of flexible estimation methods for nuisance objects.

While DML offers a framework for combining flexible nuisance estimation with valid asymptotic inference, its implementation raises important challenges. Available theoretical results assume the nuisance functions are estimated with sufficiently high accuracy. Achieving these convergence rates for modern ML methods often demands strong assumptions and special tuning, and they may not hold for off-the-shelf algorithms. These theoretical qualifications manifest in practical problems where empirical results depend on implementation choices, such as selecting and tuning an ML method for nuisance estimation. 
Assessing the quality of nuisance estimators is often difficult and,  
in some applications, different seemingly reasonable choices can lead to substantively different conclusions. 
%Further, available theoretical results assume the nuisance functions are estimated with sufficiently high accuracy, a requirement that can be challenging in practice. Achieving the necessary convergence rates for modern machine learning methods often demands strong assumptions or special tuning, and may not hold for off-the-shelf algorithms. 
%A central theme of this review is therefore not only to explain how DML works, but also to emphasize the need for transparency, robustness checks, and careful diagnostic analysis when applying these methods in practice. 
This paper therefore aims not only to motivate and explain DML, but also to guide its application in empirical research, emphasizing the need for careful diagnostic analysis and robustness checks. To this end, we divide our review into two parts.

%This paper aims to motivate and guide the application of DML in empirical research.  To this end, we divide our review into two parts. 
First, in Sections \ref{sec:DML} and \ref{subsec:DML_generic}, we introduce the DML blueprint at a high level. Section~\ref{sec:DML} discusses the practical implications of nuisance estimation and the role of DML's two key components in their remedy. Section~\ref{subsec:DML_generic} summarizes the asymptotic properties of DML and contains algorithmic details on implementation of generic DML estimators. 

In the second part of the paper, we turn to simulations and empirical applications, found in Section~\ref{sec:simulations} through Section~\ref{sec:application_monopsony}. These examples illustrate DML and provide discussion of key issues that arise in its implementation.

In Section~\ref{sec:simulations}, we present results from two simulation examples. The first is a simple linear IV example that demonstrates the importance of cross-fitting. The second compares the benefits of the DML average treatment effect estimator with inverse propensity weighted and regression-based estimators. 

In Section \ref{sec:example_hrs}, we illustrate how DML can be leveraged to reduce the dependence on functional forms in staggered adoption designs with covariates. We revisit the analysis of \citet{dobkin2018economic}, who study the economic consequences of hospital admission. We estimate group-time average treatment effects on the treated under a conditional parallel trends assumption, and show how DML inference applies to dynamic average treatment effects. We note that cross-fitting introduces an additional source of randomness induced by sample-splitting. Part of our aim in this example is to illustrate a simple approach to aid in gauging the impact of this source of randomness. 

In Section \ref{sec:application_monopsony}, we apply DML to estimation of regression coefficients in the presence of complex covariates. We specifically revisit \citet{dube2020} who apply DML to estimate the labor supply elasticity in online labor markets using textual controls. We have two main goals in this section: to illustrate the use of complex non-tabular data and, more importantly, to illustrate that DML estimates can vary substantially across otherwise reasonable choices of machine \emph{learners} (i.e., algorithms used to estimate nuisance functions).
%, sometimes leading to qualitatively different conclusions about economically meaningful parameters. In particular, this application highlights that conclusions drawn from DML can be highly sensitive to the set of learners considered, a point that becomes evident when comparing results across specifications in this example.
This sensitivity can lead to qualitatively different conclusions about economically meaningful parameters. 
%Given that different learners can lead to substantively different results and that 
Because it seems difficult to know \emph{ex ante} exactly which learner one should choose in these situations, we use this example to discuss robustness checks and suggest strategies for selecting ML algorithms.

Section \ref{sec:discussion} concludes by summarizing takeaways, raising some caveats, and pointing to potential directions for further research. 

To complement the present article, we provide additional resources on our regularly updated website \href{https://dmlguide.github.io}{\underline{dmlguide.github.io}}. The materials include replication files, additional examples with code, references to DML software packages, and links to other resources.

In terms of scope, we emphasize that DML---or any other estimation framework---cannot replace careful reasoning about economic parameters and identifying assumptions. Rather, with a well-defined target parameter and corresponding identifying assumptions, DML can aid in obtaining estimates of the target parameter in the presence of complex data structures and without relying on pre-specified functional form assumptions. In other words, DML is useful only after a target parameter is defined and the assumptions linking observed data to that parameter are well understood.\footnote{This treatment parallels \citet{heckman2007econometric}, who stress that estimation plays a limited role relative to defining a target parameter and articulating the assumptions that connect it to the data. %summarize the three distinct tasks of causal analysis as 1) definition of a target parameter, 2) identification of the parameter in a population, and 3) estimation of the parameter with data. DML contributes solely to the final task.
} 
With this understanding, we discuss identification assumptions only with the aim of illustrating the economic content of the applications. Further, we will not review specific ML methods. \citet{varian2014big}, \citet{Sendhil:MLAppliedMetrics}, \citet{athey2019_ml}, and \citet{dell2024deep} provide reviews of ML methods targeted at economists. \citet{Hastie2009} and \citet{islp} are classic textbook treatments of popular ML methods.

\bigskip

\noindent\textbf{\textit{Literature.}} 
Inference about low-dimensional target parameters in the presence of high-dimensional or nonparametric nuisance components has a long history in econometrics and statistics. Classic reviews such as \citet{newey:handbook}, \citet{yatchew1998nonparametric}, \citet{li:racine:book}, \citet{chen:handbook76}, and \citet{ichtodd:handbook74} synthesize early work on semiparametric and nonparametric methods, emphasizing how nuisance parameters can be accommodated without fully specifying the data-generating process. More recent surveys shift the focus toward the use of modern machine learning tools for nuisance estimation and their implications for inference; see, e.g., \citet{chernozhukov2018}, \citet{diaz2020machine}, \citet{hines2022demystifying}, and \citet{kennedy:review}. Our review complements these contributions by emphasizing the practical consequences of nuisance estimation choices for applied empirical work.%The problem of inference about a low-dimensional target parameter in the presence of a high-dimensional nuisance parameter has been the topic of a rich literature in econometrics and statistics. Reviews include \citet{newey:handbook}, \citet{yatchew1998nonparametric}, \citet{li:racine:book}, \citet{chen:handbook76}, \citet{ichtodd:handbook74}, and---more recently---\citet{chernozhukov2018}, \citet{diaz2020machine}, \citet{hines2022demystifying} and \citet{kennedy:review}. This paper complements these reviews of semiparametric theory with a greater focus on practical implications of nuisance estimation and its solutions in empirical research.

DML combines two ideas with deep roots in the semiparametric inference literature: Neyman orthogonal scores and sample splitting. \citet{NeymanCalpha} introduced orthogonal scores in the context of efficient parametric hypothesis testing. Orthogonal scores later played a central role in the development of modern semiparametric estimation, especially in settings with high-dimensional or nonparametric nuisance parameters. Key contributions developing these ideas include \citet{vaart:1991}, \citet{andrews94}, and \citet{newey94}, with a comprehensive textbook treatment provided by \citet{vdV}.%See, for example, \citet{vaart:1991}, \citet{andrews94}, and \citet{newey94}. A textbook treatment is given in \citet{vdV}.

Sample splitting has also played a long-standing role in semiparametric inference. It appears in several early contributions to semiparametric estimation; see, for example, \citet{hasminskii1978}, \citet{bickel:1982}, \citet{pfanzagl82book}, \citet{schick1986asymptotically}, and \citet{bickelritov1988}. In economics, sample splitting has long been used in instrumental variable estimation to mitigate bias from many instruments. See, for instance, \citet{ssiv} and \citet{AIK:JIVE} for foundational work and \citet{NeweyEtAl-JIVE}, \citet{hansen:kobzur}, and \citet{chyn2024examiner} for current developments. %More recently, sample splitting has gained renewed attention as a tool for reducing overfitting bias in high-dimensional and machine learning contexts.\footnote{See, e.g., \citet{robins2008higher}, \citet{ayyagari2010applications}, \citet{belloni2010lasso}, \citet{Belloni2012}, \citet{fan2012variance}, \citet{robins2013new}, \citet{vdl:AdaptiveTarget}, \citet{robins2017minimax}, \citet{wager2018}, \citet{athey2019e}, and \citet{athey2021a}.}

More recently, sample splitting and variations such as cross-fitting have gained renewed attention in high-dimensional contexts. %, where flexible nuisance estimators are prone to overfitting. 
A growing literature shows how these techniques can mitigate problems introduced by overfitting and improve inference when modern ML methods are used for nuisance estimation. See, for example, \citet{robins2008higher}, \citet{belloni2010lasso}, \citet{Belloni2012}, \citet{fan2012variance}, \citet{robins2013new}, \citet{vdl:AdaptiveTarget}, \citet{robins2017minimax}, \citet{wager2018}, \citet{athey2019e}, and \citet{athey2021a}.

DML is also related to targeted maximum likelihood (or minimum loss) estimation, which was introduced in \citet{SRR1999-rejoinder} for treatment effect estimation and generalized by \citet{van2006targeted}; see also \citet{vanderlaan:book}. \citet{zheng2011cross} discuss benefits of sample splitting for targeted maximum likelihood learning. \citet{diaz2020machine} expands on the difference between DML and targeted maximum likelihood estimation.

Our focus in this review is on the practical implications of nuisance estimation and the core ideas that motivate DML. Accordingly, we do not attempt a comprehensive survey of the rapidly expanding literature that extends DML in a wide range of directions. We nevertheless highlight several representative strands of this work.

A number of papers adapt DML to canonical empirical settings, including panel data and difference-in-differences designs \citep{chang2020double,chiang2022multiway, klosin:2023cont, abadie:matrix, HHZ:DMLcontDID, clarkeDoubleMachineLearning2024,chiang2025double}. Related contributions study the use of DML in instrumental variables and proxy control settings \citep{DMLDensity, deaner2021proxy, singh2024double}. A growing body of research also examines treatment effect and policy parameters, including incremental and dynamic treatment effects, nonparametric policy learning, and localized estimands that depend on complex nuisance components \citep{BMBK:incremental, Syrgkanis:DTE, klosin:admlcont, nie2021quasi, semenova2021debiased, DL:DMLnp, FS:orthogonallearning, kennedy2023, sasaki:DMLpolicy, kallus2024localized}. Other extensions address specific econometric complications. These include settings with generated regressors \citep{escanciano:LRgenerated}, partial or set identification \citep{semenova2023debiased}, and sample selection \citep{Bia:SampleSelectDML}. Complementing these application-driven contributions, several papers develop general frameworks for the automatic construction of Neyman orthogonal moments for broad classes of target parameters \citep{chernozhukov2021automatic, FLM:het, chernozhukov2022locally, chernozhukov2022automaticECMA, CNS:globalandlocal}.

%Other extensions address specific econometric complications. These include settings with generated regressors \citep{escanciano:LRgenerated}, partial or set identification \citep{semenova2023debiased}, and sample selection \citep{Bia:SampleSelectDML}. A related literature studies inference on high-dimensional or functional target parameters, where orthogonality is used to mitigate the impact of nuisance estimation even when the parameter of interest itself is complex \citep{kennedy2023, nie2021quasi, semenova2021debiased, FS:orthogonallearning}.

More broadly, DML allows researchers to avoid auxiliary parametric assumptions. While these assumptions simplify estimation, they are seldom motivated by economics and can be detrimental for applications that aim to estimate causal parameters. This perspective connects DML to the recent literature highlighting that statistically convenient estimands, often based on linear models, may fail to even approximate causal effects. Such failures have been documented in difference-in-differences settings \citep{de2020two, goodman2021difference, sun2021estimating, callaway:santanna, baker2022much, Chaisemartin:review, roth2023s, borusyak2024revisiting}. Related concerns arise in linear regression with multiple treatments \citep{goldsmith-pinkham2022} and in instrumental variables settings \citep{blandhol2022tsls}. By allowing for flexible estimation of nuisance parameters, DML provides a framework for inference that obviates the need for convenient but potentially detrimental parametric assumptions.

\section{Key Ingredients of DML
}\label{sec:DML}

This section describes the two essential components that define DML: Neyman orthogonality and cross-fitting. Together, they help control the sensitivity of the target estimator to nuisance estimation, which can substantially improve both the reliability of point estimates and the quality of conventional asymptotic approximations. By alleviating this sensitivity, DML further opens the door for researchers to use a wide range of flexible estimators, including many modern machine learners, for estimating nuisance parameters. 

%%%%%%%%%%%%%%%%%%%%%%%%%%%%%%%%%%%%%%%%%%%%%%%%%%%%%%%%%%%%%%%%%%%%%%%%%%%%%%%%%%%%%%%%%
%%%%%%%%%%%%%%%%%%%%%%%%%%%%%%%%%%%%%%%%%%%%%%%%%%%%%%%%%%%%%%%%%%%%%%%%%%%%%%%%%%%%%%%%%
\subsection{A Semiparametric Framework for DML}\label{subsec:setup}

We frame our discussion of DML within a relatively general semiparametric framework. There are two key elements of the framework. First, we have a target parameter of interest, $\theta_0$, which is low-dimensional; e.g., $\theta_0$ may be an average treatment effect or a fixed vector of regression coefficients. Second, we have a nuisance parameter $\eta_0$ which may be high-dimensional and potentially complex. In many examples, $\eta_0$ is a vector of conditional expectation functions, such as outcome regressions and propensity scores, though it may take other forms.

Throughout this review, we focus on the case where we observe an \iid sample $\{W_i : i=1,\ldots,n\}$ from a random vector $W$. Each $W_i$ collects the variables relevant for individual $i$. For example, $W_i$ might include an outcome $Y_i$, a treatment variable $D_i$, a vector of controls $X_i$, and excluded instruments $Z_i$. This structure also extends to cross-sectional and panel settings with arbitrary temporal dependence and fixed $T$.\footnote{To account for cluster dependence, one may simply redefine $W_i$ to include the data of the $i$th individual over multiple time periods---e.g., $W_i = (Y_{i,t}, D_{i,t}, X_{i,t})_{t=1}^T$.} %We discuss extensions to more general dependence structures in Remark \ref{rem: dependence}.

We assume the target parameter is identified by moment conditions
\begin{align}\label{eq: score_identification}
    \Ep\left[m(W; \theta_0, \eta_0)\right] = 0,
\end{align}
where $m(\cdot; \theta, \eta)$ is a known score function indexed by $\theta$ and nuisance parameter $\eta$ with true values $\theta_0$ and $\eta_0$. We focus exclusively on the case where the score function $m(\cdot; \theta, \eta)$ defines as many constraints as we have parameters of interest, but note that the framework extends to other settings such as GMM as discussed, e.g., in \citet{chernozhukov2018}. Throughout, we assume that the target parameter is strongly identified in the sense that \eqref{eq: score_identification} has a unique solution and satisfies regularity conditions such that $\sqrt{n}$-consistent and asymptotically normal inference for $\theta_0$ would be achievable if $\eta_0$ were known.\footnote{Extension to weakly identified settings is possible as in, e.g., \citet{chernozhukov2015} and \citet{ma:hdwiv}.} 

This semiparametric framework captures a large range of common parameters of interest in empirical research. We discuss four illustrative examples below.

\bigskip

%------------------------------------------------------------
\noindent \textbf{Example 1. Linear Regression Coefficient.}  
Consider linear regression with a single variable of interest $D$ and a $p \times 1$ vector of controls $X$ that may include a constant: % to capture an intercept:
\begin{align}\label{eq: lm}
    Y = \theta_0 D + X'\beta_0 + \varepsilon, \ \ \Ep[D \varepsilon] = 0, \ \ \Ep[X \varepsilon] = 0_p
\end{align}
where $0_p$ denotes a $p \times 1$ vector of zeros. The coefficient $\theta_0$ on $D$ is the target parameter. The vector of coefficients $\beta_0$ on controls $X$ is the nuisance parameter.

The traditional textbook approach is to estimate both $\theta_0$ and $\beta_0$ by applying OLS to equation \eqref{eq: lm}. This problem can be framed as a semiparametric estimation task by explicitly targeting $\theta_0$ separately from the nuisance parameters. In the linear regression example, this corresponds to another textbook approach: partialling out.

For any random variable $A$, let $\eta_{A,0} = \arg\min_{\eta} \Ep[(A - X'\eta)^2]$ denote the best linear predictor coefficient of $A$ given $X$. This definition implies the orthogonality condition 
\begin{align}\label{eq: lm normal}
    \Ep[X(A-X'\eta_{A,0})] = 0_p.
\end{align}
A valid score for $\theta_0$ is then
\begin{align}\label{eq: lm score}
m_{LM}(W;\theta,\eta) = \left[(Y - X'\eta_Y ) - \theta(D -  X'\eta_D)\right](D - X'\eta_D),
\end{align}
where the nuisance parameter is $\eta = (\eta_Y',\eta_D')'$ with true value $\eta_0 = (\eta_{Y,0}',\eta_{D,0}')'$. By \eqref{eq: lm} and the orthogonality condition, \eqref{eq: lm normal}, $\Ep[m_{LM}(W;\theta_0,\eta_0)] = 0$.

Equation \eqref{eq: lm score} is the population moment condition underlying the partialling out interpretation of linear least squares regression. It corresponds to projecting $Y$ and $D$ onto $X$ and then estimating $\theta_0$ from a regression using the resulting residuals.

By Frisch-Waugh-Lovell, this score yields the same estimator as OLS of $Y$ on $(D,X)$ in the low-dimensional linear setting. The value of writing the problem in this way is therefore 
primarily conceptual. 
%not computational in this special case, but conceptual. 
It makes explicit the construction of a score for the target parameter $\theta_0$ by projecting onto covariates $X$ and appropriate partialling out. This familiar approach generalizes and lies at the core of the key ``Neyman orthogonality'' property---to be discussed in Section \ref{sec: ingredients}---that is fundamental for DML. We verify Neyman orthogonality of the score \eqref{eq: lm score} in Appendix \ref{app: neyman verification}, and discuss a more general construction of Neyman orthogonal scores via ``partialling out'' in Appendix \ref{app:no_construction}.
%moment condition for identifying $\theta_0$ in way that provides insight in more general settings. Importantly, the score \eqref{eq: lm score} satisfies the key ``Neyman orthogonality'' property---to be discussed in Section \ref{sec: ingredients}---that is fundamental for DML. 

\hfill$\qed$ %Frisch–Waugh–Lovell “partialing out’’ theorem. It corresponds to regressing $X$ out from both $Y$ and $D$ and then estimating $\theta_0$ from a regression using the resulting residuals. $ $\hfill$\qed$

\bigskip

%------------------------------------------------------------
\noindent\textbf{Example 2. Partially Linear Regression Coefficient.}
As discussed in the Introduction, partially linear regression (PLR), 
\begin{align}\label{eq: plm}
    Y = \theta_0 D + g_0(X) + \varepsilon, \ \ \Ep[D \varepsilon] = \Ep[\varepsilon|X] = 0,
\end{align}
is a natural, flexible generalization of multiple linear regression.\footnote{Analogous to linear regression, PLR can be motivated in similar manner as the best \emph{partially} linear approximation to the conditional expectation function. Arguments for economic interest in the best ``fully'' linear approximation to the conditional expectation function as outlined, e.g., in \citet[Ch. 3]{angrist2009mostly}, also make PLR an attractive baseline choice in many economic analyses.} 
 
There are several moment conditions for identifying the PLR coefficient $\theta_0$. Two leading examples are based on the score functions
\begin{align}\label{eq: plm nonorth}
m_{naive}(W;\theta,\eta) &= (Y-g(X)-\theta D)D, \\
\label{eq: plm orth}
m_{PLM}(W;\theta,\eta) &= \left[(Y - \ell(X)) - \theta(D - r(X))\right](D - r(X)), 
\end{align}
where the nuisance parameters are $\eta(X)=g(X)$ in \eqref{eq: plm nonorth} with true value $g_0(X)$, and $\eta(X)=(\ell(X),r(X))$ in \eqref{eq: plm orth} with true values $\ell_0(X)=\Ep[Y|X]$ and $r_0(X)=\Ep[D|X]$. 

The first score is equivalent to regressing $Y-g(\cdot)$ against $D$. The second score corresponds to a ``partialling out'' approach where both $Y$ and $D$ are residualized with respect to $X$ before regressing the residuals on each other. The latter mirrors the Frisch–Waugh–Lovell logic from linear regression but now allows $X$ to enter flexibly. Note that \eqref{eq: plm orth} corresponds to the treatment of the partially linear model in \citet{robinson}. While both scores identify the target parameter $\theta_0$, only $m_{PLM}$ satisfies the key ``Neyman orthogonality'' property
%---to be discussed in Section \ref{sec: ingredients}---
that is fundamental for DML. We verify this in Section~\ref{sec: ingredients}. \hfill$\qed$ 

\bigskip

%------------------------------------------------------------
\noindent\textbf{Example 3. Linear IV Coefficient.}
In linear instrumental variable (IV) models, it is often unclear how best to use instruments. For instance, when rainfall or weather variables are employed as instruments, researchers face choices such as whether to use rainfall in levels, logarithms, squared terms, or deviations from historical averages \citep[e.g.,][]{hidalgo2010a,gilchrist2016something,dustmann2017}.

Formally, consider the linear structural equation $Y = \theta_0 D + \varepsilon$ where $D$ is an endogenous variable, $Z$ is a vector of excluded instruments, $\Ep[\varepsilon|Z]=0$ holds, and we abstract from other covariates. A natural score function is then
\begin{align}\label{eq: iv}
    m_{IV}(W;\theta,\eta) = (Y - \theta D)\eta(Z)
\end{align} 
where the true value of the nuisance function is $\eta_0(Z) = \Ep[D|Z]$, which corresponds to the optimal instrument under homoskedasticity. $\Ep[m_{IV}(W;\theta_0,\eta_0)] = 0$ then follows immediately from the exclusion restriction $\Ep[\varepsilon|Z] = 0$.\footnote{Under mean independence, any function $g(Z)$ serves as a valid instrument in the sense of satisfying the moment condition $\Ep[(Y-\theta_0 D)g(Z)] = 0$. However, the instrument relevance condition requires that $\Ep[g(Z) D] \neq 0$. $\Ep[g(Z) D]$ is also tightly tied to the efficiency of the IV estimator. Under homoskedasticity, the choice of $g(Z) = \eta_0(Z) = \Ep[D|Z]$ produces an asymptotically efficient estimator of $\theta_0$.} 

The IV score satisfies the key ``Neyman orthogonality'' property that is fundamental for DML. We verify this in Appendix \ref{app: neyman verification}.$ $\hfill$\qed$

\bigskip

%------------------------------------------------------------
\noindent\textbf{Example 4. Average Treatment Effect.}
A central policy-relevant parameter is the average treatment effect (ATE) of a binary treatment $D$ on an outcome $Y$ defined as 
\begin{align}\label{eq:def_ATE}
    \theta_0 = \Ep\left[Y(1)-Y(0)\right]
\end{align}
where $Y(d)$ is the potential outcome under treatment status $d\in\{0,1\}$. 

In non-experimental settings, identification of the ATE relies on two standard conditions: overlap and unconfoundedness \citep[e.g.,][]{ImbensRubin2015}. Overlap requires that the probability of treatment is bounded away from 0 and 1 across all covariate values: $0 < \text{Pr}(D = 1|X=x) < 1$ for all $x$. That is, we should see treatment and control observations at all values of $X$. Unconfoundedness requires that the treatment status is independent of potential outcomes after conditioning on the covariates: $(Y(1),Y(0)) \perp D|X.$ That is, treatment is as good as randomly assigned after conditioning on $X$.

Under these assumptions, the ATE can be identified using moment conditions. We consider two commonly used scores, the inverse propensity weighted (IPW) score and the augmented IPW (AIPW) score \citep{newey94,rrz}:
\begin{align}\label{eq: IPW}
    m_{IPW}(W; \theta, \alpha)  &=  \alpha(D, X) Y -\theta, \\ 
    \label{eq: AIPW}
    m_{AIPW}(W; \theta, \eta) &=  \alpha(D, X)(Y- \ell(D, X)) + \ell(1,X) - \ell(0,X) - \theta,
\end{align}
where $W = (Y, D, X)$. The true value of the nuisance parameters are $\alpha_0(D, X) = \frac{D}{r_0(X)} - \frac{(1-D)}{1 - r_0(X)} $, $r_0(X) = \Ep[D\vert X]$, and $\ell_0(D,X) = \Ep[Y\vert D, X]$. Under overlap and unconfoundedness, it can be shown that both $\Ep[m_{IPW}(W;\theta_0,\eta_0)] = 0$ and $\Ep[m_{AIPW}(W;\theta_0,\eta_0)] = 0$. 

Importantly, only the AIPW score---also referred to as the ``doubly robust'' score---satisfies the key ``Neyman orthogonality'' property that is fundamental for DML. The IPW score is not Neyman orthogonal and should not be used together with generic machine learners. We verify Neyman orthogonality of the AIPW score in Appendix \ref{app: neyman verification}, and illustrate empirical consequences of (non-)orthogonality in Section \ref{sec:example_401k}.$ $\hfill$\qed$ 

\bigskip

Identification of the target parameter in each of these examples depends on nuisance parameters. Outside of special cases such as randomized controlled trials where the propensity score $\Ep[D|X] = r_0(X)$ is known by design, these nuisance parameters are generally unknown and thus need to be estimated. 

In general, many moment conditions will exist for any target parameter. Examples~2 and 4 illustrate this by presenting two different moments that each identify the parameter of interest and could, in principle, be used for estimation. However, in both cases, only one of the proposed moment functions satisfies a key condition---Neyman orthogonality---that is crucial for obtaining reliable estimates in the presence of nuisance parameters.  

In the next subsection, we discuss statistical issues stemming from the estimation of these nuisance parameters. Then, in Section~\ref{sec: ingredients}, we outline how the combination of DML's two essential components alleviates the impact of nuisance estimation on the main inferential target. Fundamentally, it is this reduction of impact that allows DML to accommodate complex estimators, including ML methods, for nuisance estimation.

%%%%%%%%%%%%%%%%%%%%%%%%%%%%%%%%%%%%%%%%%%%%%%%%%%%%%%%%%%%%%%%%%%%%%%%%%%%%%%%%%%%%%%%%%
%%%%%%%%%%%%%%%%%%%%%%%%%%%%%%%%%%%%%%%%%%%%%%%%%%%%%%%%%%%%%%%%%%%%%%%%%%%%%%%%%%%%%%%%%
\subsection{Impact of Nuisance Parameter Estimation}\label{sec: expansion}

Suppose that we have at our disposal a first-step estimator, $\hat\eta$, for the nuisance parameter $\eta_0$. This might be a parametric estimator such linear regression or a flexible, nonparametric learner from the ML toolbox. A plug-in estimator for the parameter of interest can be constructed as the solution to the sample average of the scores: 
\begin{align}\label{eq: sample_analog}
    \hat{\theta}: \quad \frac{1}{n} \sum_{i=1}^n m(W_{i}; \hat{\theta}, \hat{\eta}) = 0.
\end{align}
Inference about $\theta_0$ follows from a standard asymptotic Taylor expansion of \eqref{eq: sample_analog} around the true parameters $(\theta_0, \eta_0)$:\footnote{We use a finite-dimensional expansion here to convey intuition. A formal treatment would require additional technicality to deal with cases where $\eta$ is a high- or infinite-dimensional object such as a function.}
\begin{align}
    \nonumber
	\frac{1}{n}\sum_{i=1}^n m(W_i; \hat{\theta}, \hat{\eta}) & = \frac{1}{n}\sum_{i=1}^n m(W_i; \theta_0, \eta_0) + \frac{1}{n}\sum_{i=1}^n \frac{\partial}{\partial \theta} m(W_i; \theta_0, \eta_0)(\hat{\theta} - \theta_0) \\
    \nonumber
	& \quad + \frac{1}{n}\sum_{i=1}^n\frac{\partial}{\partial \eta} m(W_i; \theta_0, \eta_0)(\hat{\eta} - \eta_0)  + \text{higher order terms} %\\
\end{align} 
\begin{align}
\begin{split}
    \label{eq: expansion1}    
    \Rightarrow \quad \sqrt{n}(\hat{\theta} - \theta_0) & = -\Bigg(\left[\frac{1}{n}\sum_{i=1}^n \frac{\partial}{\partial \theta} m(W_i; \theta_0, \eta_0)\right]^{-1}\underbrace{ \frac{1}{\sqrt{n}}\sum_{i=1}^n m(W_i; \theta_0, \eta_0)}_{\text{CLT}}  \\ %\ \\ 
    & \quad + \underbrace{\left[\frac{1}{n}\sum_{i=1}^n \frac{\partial}{\partial \theta} m(W_i; \theta_0, \eta_0)\right]^{-1}\frac{1}{\sqrt{n}}\sum_{i=1}^n\frac{\partial}{\partial \eta} m(W_i; \theta_0, \eta_0)(\hat{\eta} - \eta_0)}_{\text{($\star$): First-order impact of nuisance estimation}} \\ %\ \\
    & \quad  + \sqrt{n}\times(\text{higher order terms}) \Bigg),
\end{split}
\end{align} 
where ``higher order terms'' capture the impact of squared estimation errors $(\hat{\eta} - \eta_0)^2$ and other remainders from the linearization. Under standard regularity conditions, the term labeled CLT in \eqref{eq: expansion1} will be approximately normal by a central limit theorem. The focus of DML is addressing the term ($\star$). 

The term ($\star$) captures the first-order impact of estimating the nuisance parameter $\eta_0$. Its presence suggests the asymptotic distribution of $\hat{\theta}$ will generally depend on the asymptotic behavior of the nuisance estimator $\hat{\eta}$. That is, estimation error in $\hat{\eta}$ propagates directly into inference about $\theta_0$, making the resulting distribution of $\hat\theta$ differ from the idealized case in which $\eta_0$ is known. In situations where the nuisance parameter is low-dimensional (e.g., when assuming a linear model with few parameters), this additional uncertainty can be adequately characterized and managed through adjustments to the asymptotic variance; see, e.g., Section 6 of \citet{newey:handbook}.

However, the first-order dependence of $\hat\theta$ on $\hat\eta$ poses substantial complications in settings where the nuisance parameter is high-dimensional and $\hat{\eta}$ corresponds to a flexible estimator. The complication results because flexible estimators are often associated with non-negligible bias and variance. As a consequence, \eqref{eq: expansion1} may be dominated by the first-order term ($\star$) that involves $\hat\eta$. In general, ($\star$) \emph{diverges} due to two issues, referred to as regularization bias and overfitting bias. %The dominance of this term then generally implies two distinct sources of bias in the target parameter estimator: regularization bias and overfitting bias. Albeit for different reasons, both biases generally imply a failure of inference that does not explicitly account for estimation of $\eta_0$.

%Regularization bias refers to the direct sensitivity of the plug-in estimator to estimation errors in $\hat{\eta}$. The terminology is motivated by the fact that high-dimensional or nonparametric methods used to estimate $\eta_0$ typically rely on regularization. That is, they introduce bias to control variance. If the plug-in estimator is sensitive to the value of the nuisance parameter, this bias can then feed through to the estimator of the target parameter, and distort and complicate its asymptotic distribution.

Regularization bias refers to the fact that neither term inside the sum in $(\star)$---$\frac{\partial}{\partial \eta} m(W_i; \theta_0, \eta_0)$ and $(\hat{\eta} - \eta_0)$---is mean zero in general. As a result, $(\star)$ is $\sqrt{n}$ times a sample average of a non-mean zero quantity, which does not converge in general. The term ``regularization bias'' reflects the fact that high-dimensional or nonparametric methods used to estimate $\eta_0$ often rely on regularization. That is, they introduce bias in order to control variance, implying that $(\hat{\eta} - \eta_0)$ will not generally be mean zero in finite samples. Importantly, researchers have some control over score functions. The first key ingredient of DML---Neyman orthogonality, discussed in more detail in Section \ref{sec: ingredients}---is exactly the requirement that estimation be based on scores for which the high-dimensional analog of $\frac{\partial}{\partial \eta} m(W_i; \theta_0, \eta_0)$ is mean zero. 

%Overfitting bias arises more subtly. Because $\hat{\eta}$ is itself estimated from data, it is a random function of the data used in its estimation rather than a fixed object. As a result, it may be correlated with the observations $W_i$ that are used to construct the moment condition. This correlation typically occurs when the same dataset is used both to estimate $\hat{\eta}$ and to evaluate the sample analog of equation \eqref{eq: intro moment}. If this dependence becomes too strong, substituting $\hat{\eta}$ distorts the moment equation and thereby the plug-in estimator.\footnote{We use the term ``overfitting bias'' to describe bias in the target parameter estimator that arises due to dependence between $\hat{\eta}$ and the data used to evaluate the moment condition. We recognize that ``overfitting'' is typically associated with inflated variance in prediction contexts. Our usage emphasizes that overfitting in nuisance estimation can induce bias in the estimator of the target parameter, and avoids the more cumbersome ``bias due to overfitting'' or less evocative ``own-observation-bias.''}

Overfitting bias, also referred to as own-observation bias, arises more subtly. Because $\hat{\eta}$ is a function of the data used in its estimation, $\hat{\eta}-\eta_0$ generally depends on the observations $W_i$ that are also used to construct the sample moment condition. This dependence typically occurs when the same dataset is used both to obtain $\hat{\eta}$ and to evaluate the sample analog of equation \eqref{eq: intro moment}, though it can arise more generally in dependent data settings. As a result, the product $\frac{\partial}{\partial \eta} m(W_i; \theta_0, \eta_0)(\hat{\eta} - \eta_0)$ in ($\star$) will generally not be mean zero, even if either term in the product is mean zero when considered in isolation.

We use the term ``overfitting bias'' to describe failures in conventional inference about the target parameter that arise from statistical dependence between $(\hat{\eta} - \eta_0)$ and the observations used in the sample moment condition. We recognize that ``overfitting'' is typically associated with inflated variance in prediction contexts. Our usage emphasizes that overfitting in nuisance estimation can lead to bias in target parameter estimation, because it inflates the dependence between nuisance estimation error and the data. 

The second key ingredient of DML---cross-fitting, also discussed in Section~\ref{sec: ingredients}---addresses overfitting bias by using sample splitting to ensure (approximate) independence between the estimation error in the nuisance function and the observations used in the sample moment condition. This independence implies that the product in the numerator of  ($\star$) is mean zero whenever either term is mean zero. %This product being mean zero will then imply that ($\star$) vanishes asymptotically under regularity conditions and assuming that $(\hat{\eta} - \eta_0)$ converges sufficiently quickly. That is, the two key ingredients of DML serve to remove the first-order impact of nuisance estimation.

%Roughly speaking, regularization bias arises because the estimation error $(\hat{\eta} - \eta_0)$ is ``too large'', while overfitting bias appears because $(\hat\eta-\eta_0)$ is ``strongly correlated'' with $W_i$. Albeit for different reasons, both biases imply a failure of inference that does not explicitly account for estimation of $\eta_0$ in general.

The two main ingredients of DML, discussed in Section~\ref{sec: ingredients}, are meant to alleviate the first-order impact of nuisance estimation. Estimation of nuisance parameters also generally affects the target parameter through the ``higher order terms'' in \eqref{eq: expansion1}. A sufficient condition for these terms %are typically dominated by objects that behave like $\|\hat{\eta} - \eta_{0}\|^2$ for a suitable norm. A sufficient condition for these terms
to be asymptotically ignorable is that the nuisance parameters are estimated accurately enough such that $\sqrt{n} \|\hat{\eta} - \eta_{0}\|^2 \rightarrow_p 0$ under a suitable norm. A mean-square convergence rate faster than $n^{-1/4}$ is a commonly cited sufficient benchmark.\footnote{Because the treatment of higher-order terms in DML is similar to that in other semiparametric approaches, we do not discuss them further. For more detailed discussion, see, e.g., \citet{chernozhukov2018},  \citet{chernozhukov2022locally}, and \citet{kennedy:review}.} 

Establishing such estimation quality guarantees for modern machine learning methods is an active area of research. Available results involve a combination of assumptions on the structure of the underlying data generating process and a choice of estimator that successfully leverages that structure. A canonical example is the lasso, which achieves suitable rates when the true regression function is sparse \citep[e.g.,][]{bickel2009simultaneous,Belloni2012}. Related results for other classes of learners similarly require strong restrictions, such as smoothness, low effective dimension, or specific forms of regularization; see, for example, results for neural networks under compositional or smoothness assumptions \citep[e.g.,][]{farrell2021,schmidt2020nonparametric} and for random forests under honesty and regularity conditions \citep[e.g.,][]{wager2018,Athey2019_grf}.%Similar structure-estimator combinations also underlie results for other ML methods, including variants of random forests and neural networks.\footnote{See, e.g., \citet{bickel2009simultaneous,Belloni2012,wager2015adaptive,wager2018,oprescu2019orthogonal,Athey2019_grf,schmidt2020nonparametric,syrgkanis:2020,farrell2021, lasso:cv, CS:bootstrapCV, kueck:boosting}.}
%It is worth noting that many of the existing results do not apply to off-the-shelf implementations with default tuning choices. See, for example, \citet{CVFL:hdrandomforests}, who obtain upper bounds on rates of convergence for random forests in a high-dimensional setting that are too slow for the above condition when using tuning close to the software default.

At the same time, recent work highlights important limitations of these results. In particular, many theoretical guarantees do not apply to off-the-shelf implementations with default tuning choices. For tree-based methods, available analyses show that, absent specific tree depth and subsampling choices, pointwise polynomial convergence rates may fail or that estimators may even be pointwise inconsistent, with only slow $L^2$ consistency established in high-dimensional settings \citep[e.g.,][]{CVFL:hdrandomforests,cattaneo:trees}. More broadly, if the nuisance function $\eta_0$ is fully nonparametric and high-dimensional without additional simplifying structure, no known method is able to attain the $n^{-1/4}$ rate required for standard DML asymptotics.

In applications, it is often unclear which assumptions are credible and which estimator is appropriate, making it difficult to judge whether the required convergence conditions plausibly hold. We thus return to the topic of the choice of ML estimator in Section~\ref{sec:application_monopsony}. 

%%%%%%%%%%%%%%%%%%%%%%%%%%%%%%%%%%%%%%%%%%%%%%%%%%%%%%%%%%%%%%%%%%%%%%%%%%%%%%%%%%%%%%%%%
%%%%%%%%%%%%%%%%%%%%%%%%%%%%%%%%%%%%%%%%%%%%%%%%%%%%%%%%%%%%%%%%%%%%%%%%%%%%%%%%%%%%%%%%%

\subsection{Ingredients of DML}\label{sec: ingredients}

To reduce the dependence of estimators for $\theta_0$ on the estimation of high-dimensional nuisances $\eta_0$, DML estimators rely on two essential components---estimation based on Neyman orthogonal scores and cross-fitting---that mitigate regularization and overfitting bias. We now discuss each in turn.

\subsubsection{Neyman Orthogonality} 

The chief difficulty in using the plug-in estimator $\hat\theta$ is its dependence on the nuisance parameter estimator $\hat\eta$. \emph{Neyman orthogonality} is a local robustness property of the score function in \eqref{eq: score_identification} that decreases sensitivity of $\hat\theta$ to errors in estimating $\eta_0$. 

Formally, in addition to identifying the target parameter by satisfying \eqref{eq: score_identification}, a \emph{Neyman orthogonal score} also satisfies 
\begin{align}
\label{eq:neyman_def}
\frac{\partial}{\partial \lambda}\big\{\Ep\left[m(W; \theta_0, \eta_0 + \lambda(\eta - \eta_0))\right]\big\}\big\vert_{\lambda=0} &= 0, \quad \forall \eta \in \mathcal{T},
\end{align}
where $\eta$ denotes a candidate value for the nuisance parameter and $\lambda$ indexes the size of a local deviation away from the true $\eta_0$. Intuitively, \eqref{eq:neyman_def} requires that, when we make a small move away from $\eta_0$ in any direction $\eta-\eta_0$, the moment condition remains unchanged.\footnote{This formulation accommodates cases where $\eta_0$ is not finite-dimensional but instead a function or other complex object belonging to an abstract space $\mathcal{T}$.} In other words, small perturbations of the nuisance parameter away from the true value do not create first-order changes in the moment function. As a result, the plug-in estimator of $\theta_0$ is less sensitive to estimation error in $\hat\eta$, thereby reducing regularization bias.

Returning to the expansion in \eqref{eq: expansion1}, consider the case where the estimator $\hat{\theta}$ is based on a Neyman orthogonal score. 
If the first-step estimation error $(\hat{\eta} - \eta_0)$ were independent of the sample $\{W_i\}_{i=1}^{n}$ used to construct the score, Neyman orthogonality would imply that the second term in \eqref{eq: expansion1} vanishes,
\begin{align}\label{eq: rem vanish}
\sqrt{n}\left(\frac{1}{n}\sum_{i=1}^n\frac{\partial}{\partial \eta} m(W_i; \theta_0, \eta_0)(\hat{\eta} - \eta_0)\right) \approx 0,
\end{align}
under the convergence requirements on $\hat\eta$ cited above and additional mild regularity conditions. Loosely, condition  \eqref{eq:neyman_def} means that $\frac{1}{\sqrt{n}}\sum_{i=1}^n\frac{\partial}{\partial \eta} m(W_i; \theta_0, \eta_0)$ behaves like a mean~0 normalized sum, which converges by a central limit theorem. As long as $(\hat{\eta} - \eta_0)$ converges to 0, the whole term disappears. %\footnote{The convergence rate conditions sufficient for ignoring higher-order terms in Section~\ref{sec: expansion} are also sufficient here.} 
Maintaining the assumption that higher order terms vanish, it follows that the asymptotic distribution of the plug-in estimator $\hat\theta$ does not depend on the nuisance estimator $\hat\eta$. Thus, inference about $\theta_0$ using $\hat{\theta}$ can proceed as if $\eta_0$ were known. This heuristic derivation can be formalized and underlies the crucial importance of Neyman orthogonality in settings with complex nuisance parameters.

An important practical implication is that estimators based on Neyman orthogonal scores yield inference about $\theta_0$ that does not depend on the detailed statistical properties of the nuisance estimator $\hat\eta$. This robustness is useful even in classical low-dimensional settings, where it avoids cumbersome variance adjustments or computationally costly resampling approaches to account for estimation of $\hat\eta$. It becomes particularly important when modern flexible methods are used, as the statistical properties of these methods are still under active development. For example, only coarse rates of convergence are currently available for many promising machine learning methods. By alleviating the first-order impact of nuisance estimation, Neyman orthogonality makes it possible to combine such methods with standard asymptotic approximations, enabling formally valid inference for $\theta_0$ while exploiting modern flexible estimators for nuisance parameters. 

We illustrate the importance of using Neyman orthogonal scores for obtaining reliable finite-sample inference in Section~\ref{sec:simulations}. There, we present simulation results demonstrating that estimators of the average treatment effect (ATE) based on non-orthogonal scores may exhibit substantial bias, and the associated confidence intervals are often severely distorted. In contrast, DML estimators, which incorporate Neyman orthogonal scores as a core component, are approximately unbiased, and their confidence intervals achieve coverage rates close to the nominal level.

\paragraph{Neyman orthogonal scores for common parameters.}
Neyman orthogonal scores are well-known and readily available for common target parameters. We present Neyman orthogonal scores for six illustrative targets in Table~\ref{tab:overview}. In Panels~(i)-(iv), we present Neyman orthogonal scores corresponding to the target parameters in Examples~1-4 from Section \ref{subsec:setup}. %\footnote{We make a slight deviation in that Panel~(i) presents a Neyman orthogonal score for the linear model with multiple control variables while Example~1 includes just one control variable.} 
In Panel~(v), we present an orthogonal score for the coefficient $\theta_0$ in the partially linear IV model
\begin{align*}
Y = \theta_0 D + g_{0}(X) + \varepsilon, \ \ \ \Ep[Z\varepsilon] = \Ep[\varepsilon|X] = 0, 
\end{align*}
where $Z$ is an excluded scalar instrumental variable. This model differs from the simple IV model considered in Example~2 in two key respects: We allow for the presence of controls, and we do not impose full mean independence of the instrument from structural unobservables. Finally, in Panel~(vi), we consider a more exotic target parameter---an average derivative corresponding to a continuous variable of interest. We present this case both because continuous variables are practically relevant and, more importantly, to highlight that nuisance parameters are not always conditional expectation functions (or projection coefficients as in the linear regression example). 

\begin{table}[t!]

\centering\footnotesize

\resizebox{\linewidth}{!}{\begin{tabular}{p{.23\linewidth}p{.29\linewidth}p{.43\linewidth}}

\toprule\toprule

\multicolumn{1}{l}{\emph{Observed Variables} ($W$)} & \multicolumn{1}{l}{\emph{Nuisance parameters} ($\eta$)} & \multicolumn{1}{l}{\emph{Neyman orthogonal score} ($m(W; \theta, \eta)$)} \\

\midrule

\\[-.3cm]

%%% LM 

\multicolumn{3}{l}{(i) {\it Target Parameter: Linear regression coefficient}}\\

\\[-.2cm]

$Y$: outcome & $\eta_{Y,0} = \arg\min_{\eta} \Ep[(Y - X'\eta)^2]$ &
\multirow{3}{=}{\centering $\left[(Y - X'\eta_{Y}) - \theta(D - X'\eta_D)\right](D-X'\eta_D)$ } \\
$D$: treatment & $\eta_{D,0} = \arg\min_{\eta} \Ep[(D - X'\eta)^2]$ & \\
$X$: controls & & \\

\\

%%% PLM 

\multicolumn{3}{l}{(ii) {\it Target Parameter: Partially linear regression coefficient}}\\

\\[-.2cm]

$Y$: outcome & $\ell_0(X) = \Ep[Y \vert X]$ &
\multirow{3}{=}{\centering $\left[(Y - \ell(X)) - \theta(D - r(X))\right](D-r(X))$ } \\
$D$: treatment & $r_0(X) = \Ep[D \vert X]$ & \\
$X$: controls & & \\

\\

%%%% Linear IV

\multicolumn{3}{l}{(iii) {\it Target Parameter: Linear IV coefficient (No controls)}}\\

\\[-.2cm]

$Y$: outcome & & \\
$D$: treatment & $r_0(Z) = \Ep[D|Z]$ & \makebox[\linewidth]{\centering $\left(Y - \theta D\right) r(Z)$} \\
$Z$: instruments & & \\

\\

%%% ATE

\multicolumn{3}{l}{(iv) {\it Target Parameter: Average treatment effect $\left(\Ep\left[\Ep[Y|D=1,X] - \Ep[Y|D=0,X]\right]\right)$}}\\

\\[-.2cm]

$Y$: outcome &
$\ell_0(D,X) = \Ep[Y \mid D, X]$ &
\multirow{3}{=}{\centering $\alpha(D, X)(Y - \ell(D, X)) + \ell(1,X) - \ell(0,X) - \theta$} \\
$D$: binary treatment &
$\alpha_0(D,X) = \frac{D}{r_0(X)} - \frac{1-D}{1 - r_0(X)}$  &
\\
$X$: controls &
\ \ \ for $r_0(X) = \Ep[D \mid X]$ &
\\

\\

%%%% PLM-IV

\multicolumn{3}{l}{(v) {\it Target Parameter: Partially linear regression coefficient with excluded instruments}}\\

\\[-.2cm]

$Y$: outcome & $\ell_0(X) = E[Y \vert X]$ & \multirow{4}{=}{\centering  $\left[(Y - \ell(X)) - \theta(D - r(X))\right](Z-h(X))$} \\
$D$: treatment & $r_0(X) = \Ep[D|X]$ & \\ 
$Z$: instrument & $h_0(X) = \Ep[Z|X]$ & \\
$X$: controls & & \\

\\

%%% Average derivative

\multicolumn{3}{l}{(vi) {\it Target Parameter: Average structural derivative}\, $\left(\Ep\left[\frac{\partial}{\partial d}\Ep[Y|D=d,X]\mid_{d=D}\right]\right)$}
\\

\\[-.2cm]

$Y$: outcome & $\ell_0(D,X) = \Ep[Y\vert D, X]$ 
& \multirow{3}{=}{\centering $\frac{\partial}{\partial d} \ell_0(d,X)\mid_{d=D} + \alpha(D, X)(Y - \ell(D, X)) - \theta$} \\
$D$: continuous treat. & $\alpha_0(D, X)=$ & \\
$X$: controls & $\qquad -\frac{\partial}{\partial d}\log f_0(d, X)\mid_{d=D}$ & \\

\\[-.3cm]

\bottomrule\bottomrule

\end{tabular}}\par\medskip
\parbox{\linewidth}{\scriptsize{\it Notes:} For each target parameter, the table lists the observed variables, nuisance parameters, and a Neyman orthogonal score function that identifies the parameter. All expectations are taken conditional on the relevant covariates (e.g., $X$ or $Z$). In panel (vi), $f_0(D, X)$ denotes the conditional density of $D$ given $X$.}

\caption{Overview of common target parameters and Neyman orthogonal scores}
\label{tab:overview}
\end{table} 

\paragraph{Deriving and verifying Neyman orthogonal scores.}
Given the importance of Neyman orthogonal scores and their relevance for DML, we provide an outline of a general structure for obtaining Neyman orthogonal scores in Appendix~\ref{app:no_construction}. %We place this discussion in an appendix due to its technical nature.
%Importantly, this accommodates settings in which the components of the orthogonal score are not available in closed form but are instead specified implicitly. This flexibility allows DML to be applied even when explicit expressions for orthogonal scores are unavailable or difficult to derive or use. See, for example, \citet{chernozhukov2022locally, chernozhukov2022automaticECMA, CNS:globalandlocal, hirschberg:wager}.

For a given score, one can generally verify Neyman orthogonality, or a lack thereof, by direct application of its definition in \eqref{eq:neyman_def}. We illustrate such a derivation for partially linear regression (Example~2) below. Similar derivations for the scores presented in Examples 1, 3, and 4 are provided in Appendix~\ref{app: neyman verification}.

\bigskip

\noindent\textbf{Example 2 (continued). Neyman Orthogonality of Partially Linear Regression Scores.}
We introduced two score functions for identifying the partially linear regression coefficient $\theta_0$, \eqref{eq: plm nonorth} and \eqref{eq: plm orth}. The first does not satisfy Neyman orthogonality, while the second---which corresponds to flexible partialling out---does.

Consider \eqref{eq: plm nonorth}, which has nuisance parameter $g(X)$. Let $\Delta g(X) = g(X)-g_0(X)$; then
\begin{align*}
    \frac{\partial}{\partial \lambda}\Ep[m_{naive}(W;\theta_0,g_0(X) + \lambda\Delta g(X))]\big\vert_{\lambda = 0} = \Ep\left[-\Delta g(X) D\right],
\end{align*}
which is generally non-zero when $D$ and $X$ are related. Hence, the score is not orthogonal.

Turning to \eqref{eq: plm orth}, we have nuisance parameters $\eta(X) = (\ell(X),r(X))$ with true value $\eta_0(X) = (\ell_0(X) = \Ep[Y|X],r_0(X) = \Ep[D|X])$. Letting $\Delta\eta(X) = \eta(X) - \eta_0(X) = (\Delta \ell(X) = \ell(X) - \ell_0(X),\Delta r(X) = r(X)-r_0(X))$, we have
\begin{align*}
    \frac{\partial}{\partial \lambda}&\Ep[m_{PLM}(W;\theta_0,\eta_0(X) + \lambda\Delta \eta(X))]\big\vert_{\lambda = 0} \\ &= \Ep\left[-\Delta \ell(X)(D-r_0(X)) - \Delta r(X)(Y-\ell_0(X)) + 2\theta_0\Delta r(X)(D-r_0(X))\right]  = 0
\end{align*}
where the last equality follows from $\ell_0(X) = \Ep[Y|X]$ and $r_0(X) = \Ep[D|X].$  $ $\hfill$\qed$

\bigskip

The intuition for Neyman orthogonality in Example~2 is instructive and corresponds to common intuition provided for partialling out. The orthogonal score uses only variation in $D$ and $Y$ that is (mean) independent of $X$, thereby isolating the identifying variation. As a result, small errors in one nuisance function can be offset by the other, making estimation more robust. In contrast, the non-orthogonal score uses all the variation in $D$ but adjusts only for the effect of controls on $Y$. Any mistakes in estimating $g(X)$ that are correlated with $D$ then directly bias the estimate of $\theta_0$, much like in a classic omitted variable scenario. Appendix \ref{app:no_construction} shows that the same partialling out intuition generalizes to average treatment effects and many other canonical target parameters.

%In the special case of linear regression (Example~1), where $g_0(X)=X'\beta_0$, $m_{naive}$ will be orthogonal if we use the OLS estimates $g(X)=X'\hat{\beta}_{OLS}$ from applying OLS to \eqref{eq: lm}. This is because OLS is unbiased and $\hat{\beta}_{OLS}-\beta_0$ is mean zero. It is when we leave this special setting and want to estimate the nuisance parameters flexibly that regularization bias emerges and estimation relying on $m_{naive}$ becomes problematic.

%%%%%%%%%%%%%%%%%%%%%%%%%%%%%%%%%%%%%%%%%%%%%%%%%%%%%%%%%%%%%%%%%%%%%%%%%%%%%%%%%%%%%%%%%
%%%%%%%%%%%%%%%%%%%%%%%%%%%%%%%%%%%%%%%%%%%%%%%%%%%%%%%%%%%%%%%%%%%%%%%%%%%%%%%%%%%%%%%%%
\subsubsection{Cross-fitting}

Overfitting bias arises due to statistical dependence between the error in the nuisance parameter estimator and the data used in constructing the plug-in estimator. As discussed above, ignoring the first-step estimation of $\hat\eta$ is justified only if the term~($\star$) in \eqref{eq: expansion1} is asymptotically negligible. In the previous section, we outlined an argument for this term vanishing that depends on Neyman orthogonality \emph{and} independence between the estimation error in the nuisance function, $\hat{\eta} - \eta_0$, and the data used to construct the sample moment condition $\{W_i\}_{i=1}^{n}$. If $\hat{\eta}$ is constructed using these same observations, the independence assumption will be violated. More generally, Neyman orthogonality alone is not sufficient to guarantee that first-step estimation of nuisance parameters can be ignored in inference about low-dimensional target parameters. To address this issue, DML relies on a second key ingredient: cross-fitting.

Cross-fitting is a form of repeated sample splitting intended to reduce the dependence between first-step estimation error and the data used to estimate the target parameter. Intuitively, if two independent datasets were available, one could be used to estimate the nuisance function $\hat{\eta}$ and the other to estimate $\theta_0$ by plugging in $\hat{\eta}$. In that case, independence between $\hat\eta - \eta_0$ and $\frac{\partial}{\partial \eta} m(W_i; \theta_0, \eta_0)$ would follow by construction. With a single sample of independent observations, we can mimic this logic by randomly splitting the sample into two partitions, or ``folds,'' using one fold to estimate $\hat{\eta}$ and the other to evaluate the score function and estimate $\theta_0$. Of course, such an approach inefficiently uses the available data because both $\eta_0$ and $\theta_0$ are estimated using only subsets of the data. Cross-fitting restores asymptotic efficiency by rotating which folds are used for each task, so all observations contribute to both steps.

The cross-fit version of the plug-in estimator with a generic score defined in \eqref{eq: sample_analog} is 
\begin{align}\label{eq: sample_analog_cf}
    \hat{\theta}^{\,\times \text{-fit}}: \quad \frac{1}{n}  \sum_{k=1}^K  \sum_{i\in I_k} m(W_{i}; \hat{\theta}^{\times \text{-fit}}, \hat{\eta}_{-k}) = 0.
\end{align}
Here, $\{I_k\}_{k=1}^K$ is a random partition of the sample of units $\{1, \ldots, n\}$ into $K$ subsamples of approximately equal size, and $\hat{\eta}_{-k}$ denotes a first-step nuisance parameter estimate constructed using only observations \textit{excluding} those in the subsample $I_k$. Because $\hat{\eta}_{-k}$ uses only observations \emph{not} in subsample $k$, the estimation error in $\hat{\eta}_{-k}$ is independent of observations in subsample $k$, which alleviates overfitting bias.
By rotating (``crossing'') samples for the estimation of nuisance parameters and the estimation of target parameters, cross-fitting asymptotically avoids losing efficiency relative to the hypothetical full-sample estimator that makes use of the true values of the nuisance parameters. 

As a byproduct that comes at no additional computational cost, cross-fitting produces out-of-sample prediction errors associated with the nuisance functions. These enable calculation of diagnostics for evaluating the choice and specification of the nuisance function estimator. For example, in partially linear regression (Example~2), we have $\eta_0=(\ell_0,r_0)$ with $\ell_0(X)=E[Y\vert X]$ and $r_0(X)=E[D\vert X]$. The \emph{cross-fitted} errors $(Y_i - \hat{\ell}_{-k}(X_i))$ and $(D_i - \hat{r}_{-k}(X_i))$---where $\hat{\ell}_{-k}$ and $\hat{r}_{-k}$ are estimated using data not containing observation $i$---are equivalent to \emph{cross-validated} errors from $K$-fold cross-validation. Their availability allows calculating metrics for learner performance such as mean-squared prediction error (MSPE) and out-of-sample $R^2$. We discuss these and other methods for evaluating the nuisance function estimators in Section~\ref{sec:application_monopsony}.

The simulation results in Section~\ref{sec:simulations} provide finite-sample evidence that cross-fitting plays an important role in delivering reliable inference. The results illustrate that cross-fitting while using non-orthogonal scores delivers relatively little benefit in the considered examples. Similarly, the results show that inference based on Neyman orthogonal scores \emph{without} cross-fitting often fails to deliver reliable inferential results. In contrast, DML estimators, which incorporate both Neyman orthogonal scores and cross-fitting, deliver the most robust performance across the considered examples.

\bigskip

\begin{remark}[Cross-fitting with Dependence]\label{rem: dependence}
Cross-fitting based on random partitions of $\{i: 1,\ldots, n\}$ applies to cross-sectional and fixed-$T$ panel settings with independence across $i$ and arbitrary temporal dependence. In panel settings, the sample is partitioned by cross-sectional units, preserving the full time series per unit. Cross-fitting can also be extended to more complex dependence structures. For example, \citet{chiang2022multiway} extend DML to settings with multiway clustered dependence, and \citet{semenova2023inference} and \citet{BN:DMLforIRF} discuss its application in time series and dynamic panels with weak dependence.
\end{remark}

\begin{remark}[Alternatives to Cross-fitting]\label{rem: no xfit}

Under special conditions on the structure of the data, overfitting can be avoided by carefully tailoring nuisance estimators, bypassing the need for cross-fitting. 
%An alternative to using cross-fitting arises if special conditions hold that imply that overfitting will not occur. In such cases, one can use methods that are carefully tailored to avoid overfitting under those conditions. 
This approach is taken in, for example, \citet{Belloni2012}, \citet{BelloniChernozhukovHansen2011}, \citet{VdG}, \citet{Montanari},  \citet{ZZ}, \citet{belloni2017program},  \citet{farrell2021}, and \citet{wiemann2026optimal}. Such results are available for relatively few ML tools. Simulations also suggest that DML with flexible learners can perform similarly to these procedures when their conditions hold and outperforms them otherwise \citep[e.g.,][]{Ahrens2024_applied}.
\end{remark}

%%%%%%%%%%%%%%%%%%%%%%%%%%%%%%%%%%%%%%%%%%%%%%%%%%%%%%%%%%%%%%%%%%%%%%%%%%%%%%%%%%%%%%%%%
%%%%%%%%%%%%%%%%%%%%%%%%%%%%%%%%%%%%%%%%%%%%%%%%%%%%%%%%%%%%%%%%%%%%%%%%%%%%%%%%%%%%%%%%%
\section{Estimation and Inference with DML}\label{subsec:DML_generic}

In this section, we define the DML estimator, present an implementation algorithm, and summarize its asymptotic properties. We remain in the general semiparametric setting of Section~\ref{sec:DML}. That is, our focus is estimation and inference for a low-dimensional target parameter $\theta_0$, which is defined by moment conditions~\eqref{eq: score_identification} and depends on an unknown (potentially high-dimensional) nuisance parameter $\eta_0$.

Within this general setting, a DML estimator is a plug-in estimator that combines \emph{both} Neyman orthogonal scores and cross-fitting.
The DML estimator of  $\theta_0$ is then 
\begin{align}\label{eq:dml_definition}
    \hat{\theta}_{DML}: \quad \frac{1}{n}  \sum_{k=1}^K  \sum_{i\in I_k}m(W_{i}; \hat{\theta}_{DML}, \hat{\eta}_{-k}) = 0,
\end{align}
where $m$ is a Neyman orthogonal score (i.e., a score that satisfies \eqref{eq: score_identification} and \eqref{eq:neyman_def}),  $\{I_k\}_{k=1}^K$ is a random partition of the sample of individuals $\{1, \ldots, n\}$ into $K$ subsamples of approximately equal size, and $\{\hat{\eta}_{-k}\}_{k=1}^K$ are cross-fitted nuisance parameter estimators.

\citet{chernozhukov2018} provide conditions that allow researchers to make valid inferential statements about $\theta_0$ using the DML estimator $\hat{\theta}_{DML}$. These conditions involve conventional sampling and regularity conditions along with the assumption that the nuisance parameter estimator, $\hat\eta$, converges sufficiently quickly, as discussed in Section \ref{sec: expansion}. Under these assumptions, the DML estimator is asymptotically normal:
$$\sqrt{n}\hat{\Sigma}^{-1/2}(\hat{\theta}_{DML} - \theta_0) \overset{d}{\to} \mathcal{N}(0, \mathrm{I}),$$ 
where 
\begin{align}\label{eq:Sigma_hat}
\begin{aligned}
\hat{\Sigma} =& \hat{J}^{-1}\left(\frac{1}{n}\sum_{k=1}^K\sum_{i\in I_k}m(W_{i}; \hat{\theta}_{DML}, \hat{\eta}_{-k})m(W_{i}; \hat{\theta}_{DML}, \hat{\eta}_{-k})^\top\right)\hat{J}^{-1 \top}, \\
\hat{J}=& \frac{1}{n}\sum_{k=1}^K\sum_{i\in I_k} \frac{\partial}{\partial \theta} \left\{m(W_{i}; \theta, \hat{\eta}_{-k})\right\}\big\vert_{\theta =\hat{\theta}_{DML}} \: .
\end{aligned}
\end{align}
Standard errors for $\hat{\theta}$ are thus given by square-roots of the diagonal values of $\hat{\Sigma}/n$.

%\paragraph{Calculation of standard errors.}
A key practical implication of using both Neyman orthogonal scores and cross-fitting is that standard errors for $\hat{\theta}_{DML}$ can be computed as if the nuisance functions were known. This approximation result holds in a variety of settings, including standard cross-sectional data, clustered cross-sectional data, and panel data with fixed $T$. In cross-sectional applications, the standard errors coincide with conventional heteroskedasticity-robust formulas. In clustered or panel settings, they are analogous to clustered standard errors that allow for arbitrary dependence across time within units $i$.

\bigskip

\begin{remark}[Semiparametric Efficiency of DML Estimators]
The DML estimator $\hat\theta_{DML}$ is semiparametrically efficient when the orthogonal score coincides with the efficient influence function of the target parameter. Methods for deriving such scores are well-established; see, for example, \citet{newey94, chernozhukov2022locally}. %We outline a basic approach in Appendix~\ref{app:no_construction}. 
%\thomas{This continues to confuse me. Appendix B does not derive efficient influence functions. Standard counterexample is PLR w/ heteroskedasticity. It also seems to me that newey94 does not claim that the derived scores are generally efficient. But I also admit that I don't understand why the scores constructed in chernozhukov2022locally are efficient scores for $\theta_0$ (or whether that is even claimed).}
%\thomas{I think I understand it now: newey94, chernozhukov2022locally etc construct scores that are efficient for $\Ep[m(W; \theta_0, \gamma_0)]$, not $\theta_0$! Pretty obvious when you think about how the projection in \eqref{eq:alpha_proj} is constructed. Of course, the two are equivalent whenever we can write $m(W; \theta_0, \gamma_0) = \tilde{m}(W;  \gamma_0) - \theta_0$ which is the case for the ATE and ATT but not for the PLR. Maybe worth a remark in the appendix? Not sure we want to discuss newey90 though.}

As examples, the orthogonal scores in Examples 1-3 correspond to the efficient influence functions for their respective target parameters under homoskedasticity. The scores for the ATE in Example 4 and for the group-time average treatment effect on the treated in Section~\ref{sec:example_hrs} are likewise efficient for their parameters. Thus, the resulting estimators are semiparametrically efficient.
\end{remark}

%\paragraph{Algorithm.}
Algorithm \ref{alg: DML} illustrates computation of $\hat{\theta}_{DML}$ in an \iid setting. Implementation proceeds in three parts. First, data are randomly split into $K$ subsamples (Step~\ref{alg_line:partition}). Second, the cross-fitted nuisance estimates are computed in each subsample (Steps~\ref{alg_line:for_start}--\ref{alg_line:for_end}). Third, estimation and inference about the target parameter takes place (Steps \ref{alg_line: DML_estimator}--\ref{alg_line:DML_se}).

\begin{algorithm} 
\caption{DML Estimation and Inference \\ \ }\label{alg: DML}
    \begin{algorithmic}[1]
    \REQUIRE A sample $\{W_{i}\}$ for $i \in \{1, \ldots, n\}$, a Neyman orthogonal score $m$, a nuisance parameter estimator $\hat{\eta}$, an integer $K$ for the number of cross-fitting folds.
    \STATE  Randomly split the sample of indices $\{1, \ldots, n\}$ into $K$ partitions $(I_k)_{k=1}^K$ of approximately equal size.  \label{alg_line:partition}
        \FOR{k = $1$ to $K$} \label{alg_line:for_start}
            \STATE Compute the nuisance parameter estimator on samples $I^c_k = \{1, \ldots, n\} \setminus I_k$: $$
            \hat{\eta}_{-k} = \hat{\eta}(\{W_{i}\}_{i\in I^c_k}).$$
        \ENDFOR \label{alg_line:for_end}
    \STATE Construct the DML estimator $\hat{\theta}_{DML}$ for $\theta_0$ as the solution to \eqref{eq:dml_definition}.     \label{alg_line: DML_estimator}
    \STATE Estimate the covariance matrix $\hat{\Sigma}$ via \eqref{eq:Sigma_hat}.\label{alg_line:DML_se}
%    \STATE Construct confidence intervals about $\theta_0$ via \eqref{eq:conf_interval}.\label{alg_line:DML_ci}
    \end{algorithmic}
\end{algorithm}

Importantly, the algorithm is highly general: it is a blueprint for estimating and performing inference on a broad range of target parameters $\theta_0$. The procedure accommodates a wide range of nuisance parameter estimators $\hat{\eta}$, including classical approaches as well as modern and emerging ML methods. 

Implementing Algorithm~\ref{alg: DML} requires researchers to make several design choices. These include selecting an appropriate scheme for generating the cross-fitting folds, choosing the number of cross-fitting folds, and specifying the nuisance function estimator. We discuss these and other implementation choices in the following sections. In Section~\ref{sec:discussion}, we provide a discussion of current best practices for the implementation of DML estimators.

\section{Two Simulation Illustrations}\label{sec:simulations}

This section employs two simulation exercises to illustrate the consequences of relying on non-Neyman orthogonal scores and of failing to perform cross-fitting when using flexible nuisance estimators such as ML methods. We discuss two simulations to underscore that bias dominates in some contexts, while regularization bias prevails in others%, in some contexts, overfitting bias exerts a stronger impact, while in others regularization bias prevails. 

The first simulation focuses on the role of cross-fitting in the linear IV setting with many instruments. 
This problem has been thoroughly studied since the mid-1990s \citep[e.g.,][]{bekker}. 
One of the recommendations to emerge from this literature is to use sample-splitting to reduce overfitting bias (often termed ``many instruments bias'' in this context). %Note that DML with a linear first stage regression is effectively split-sample IV \citep[henceforth SSIV;][]{AngristKruegerSplitSample1995,AIK:JIVE}.\footnote{More precisely, \citet{AngristKruegerSplitSample1995} suggest splitting the sample randomly into two folds, using the first fold to estimate the IV first stage via linear regression, and the second to estimate the target parameter. That is, they do not cross-fit, though note the possibility of doing so in their conclusion. Furthermore, DML with a linear learner and $K=n$ is exactly jackknife IV \citep{AIK:JIVE}.} One can thus view DML in this context as an extension of split-sample IV that accommodates the use of flexible learners for estimating the first stage. 
Our simulation revisits this classical many instrument setup by comparing linear IV estimation with ML-based IV estimation. As each of these approaches relies on Neyman orthogonal scores (see Example~3), the IV setting isolates the impact of overfitting. We show that cross-fitting alleviates overfitting bias and that using first stage estimators other than OLS can lead to efficiency gains.

Our second example turns to the estimation of average treatment effects, where we emphasize the role of Neyman orthogonality in delivering valid inference. Using a calibrated simulation, we compare estimators based on orthogonal and non-orthogonal scores, with and without cross-fitting. The results highlight that only the combination of orthogonality and cross-fitting yields estimators with reliable sampling behavior, thereby illustrating why these two ingredients are central to the DML framework.

\subsection{Instrumental Variables with Many Instruments}\label{subsec:example_weakiv}

We present simulation results from the canonical many-instrument linear IV model (see Example~3). We generate data as \iid realizations from
\begin{align*}
    Y = \theta_0 D + \varepsilon, \qquad 
    D = g_0(Z) + \nu, \qquad (\varepsilon,\nu)' \perp Z.
\end{align*}
Here $Z$ are $p=200$ simulated instruments, $\theta_0=0$ is the parameter of interest, and $(\varepsilon,\nu)'$ are correlated error terms drawn from the normal distribution. We define the nuisance function as $g_0(Z) = Z'\pi_0$, where the first six elements of $\pi_0$ are set to 0.1 and the remaining entries are set to 0. That is, only the first six instruments carry any signal. In our setting, OLS suffers from an upward bias of around 0.5. All results are based on a sample size of $n=1000$ and 1000 simulation replications.\footnote{Specifically, we generate $Z \sim N(0_p, .4 \mathrm{I}_p + .6 \iota_p \iota_p')$ and $(\varepsilon,\nu)'=N(0_2,\Sigma)$ with $\Sigma_{11}=\Sigma_{22}=1$ and $\Sigma_{21}=\Sigma_{21}=.6$. We use $0_p$ and $\iota_p$ to denote a $p \times 1$ vector of zeros and ones, respectively; $\mathrm{I}_p$ denotes a $p \times p$ identity matrix.}

\begin{comment}
% OLD VERSION:
We present simulation results from the canonical linear IV model with many instruments (see Example~3). Specifically, we generate data as \iid realizations from
\begin{align*}
    Y = \theta_0 D + \varepsilon, \qquad 
    D = g_0(Z) + \nu, 
\end{align*}
with $(\varepsilon,\nu) \perp Z$ and
\begin{align*}
    Z  \sim N(0_p, .4 \mathrm{I}_p + .6 \iota_p \iota_p'), \qquad
    \left(\begin{array}{c} \varepsilon \\ \nu \end{array}\right)  \sim N\left(\left(\begin{array}{c} 0 \\ 0 \end{array}\right),\left(\begin{array}{cc} 1 & .6 \\ .6 & 1 \end{array}\right)\right), 
\end{align*}
where $0_p$ denotes a $p \times 1$ vector of zeros, $\iota_p$ denotes a $p \times 1$ vector of ones, and $\mathrm{I}_p$ denotes a $p \times p$ identity matrix. We set the sample size to $n=1000$ and the number of instruments to $p=200$. We define the nuisance function as $g_0(Z) = Z'\pi_0$ with $\pi_0 = 0.1\cdot(\iota_6',0_{194}')'$, implying that the first stage relationship corresponds to a sparse linear model. Finally, we set the parameter of interest to $\theta_0 = 0$. Use of OLS (i.e., ignoring endogeneity) would result in an upward bias in the estimated coefficient of about 0.5. All results are based on 1000 simulation replications.
\end{comment}

We compare five estimators, each based on a Neyman orthogonal score. As a benchmark, we report results from the oracle IV estimator that knows the true first-stage function and uses $Z'\pi_0$ as the instrument (Oracle IV). This estimator is infeasible in practice but serves as an efficient baseline. We next consider two-stage least squares using all 200 instruments (2SLS). As is well-known from the many-instruments literature, 2SLS performs poorly when the number of instruments is large. We implement DML with a linear first stage regression using all 200 instruments. 
Note that this DML estimator is essentially split-sample IV \citep[SSIV;][]{AngristKruegerSplitSample1995,AIK:JIVE}.\footnote{More precisely, \citet{AngristKruegerSplitSample1995} suggest splitting the sample randomly into two folds, using the first fold to estimate the IV first stage via linear regression, and the second to estimate the target parameter. That is, they do not cross-fit, though note the possibility of doing so in their conclusion. Furthermore, DML with a linear (first-stage) regression and $K=n$ is exactly jackknife IV \citep{AIK:JIVE}.} One can thus view DML in this context as an extension of SSIV that also accommodates the use of flexible first stage estimators, for example, 
%As noted above, this approach is closely related to the SSIV estimator of \citet{AngristKruegerSplitSample1995}. %To mimic a situation where the researcher is unsure of the correct functional form or which instruments matter, we employ gradient boosted trees to estimate the first-stage nuisance function (Boosted Trees). This version does not use cross-fitting. Finally, we apply the full DML procedure using gradient boosted trees (DML with Boosted Trees).
%To mimic a situation where research 
because the researcher does not want assume linearity and is unsure about which instruments matter. To this end, we also estimate the first stage using gradient boosted trees, both without cross-fitting (Boosted Trees) and with cross-fitting (DML with Boosted Trees). While one should consider alternative learner choices in practice, %as we discuss in Section~\ref{sec:application_monopsony}, 
we focus on gradient boosted trees with a maximum tree depth of four, a learning rate of 0.1, and early stopping based on a 20\% validation set. We use 5-fold cross-fitting for SSIV and DML with gradient boosted trees.

%We use 5-fold cross-fitting for SSIV and DML with gradient boosted trees. For this illustration, we focus on gradient boosted trees with a maximum tree depth of four, a learning rate of 0.1, and early stopping based on a 20\% validation set. We defer discussion of learner choice and specification to Section~\ref{sec:application_monopsony}.

\begin{figure}[!t] 
\caption{Histograms of IV Coefficient Estimates}\label{fig:iv_sim}
    \centering
    \includegraphics[width=0.9\textwidth]{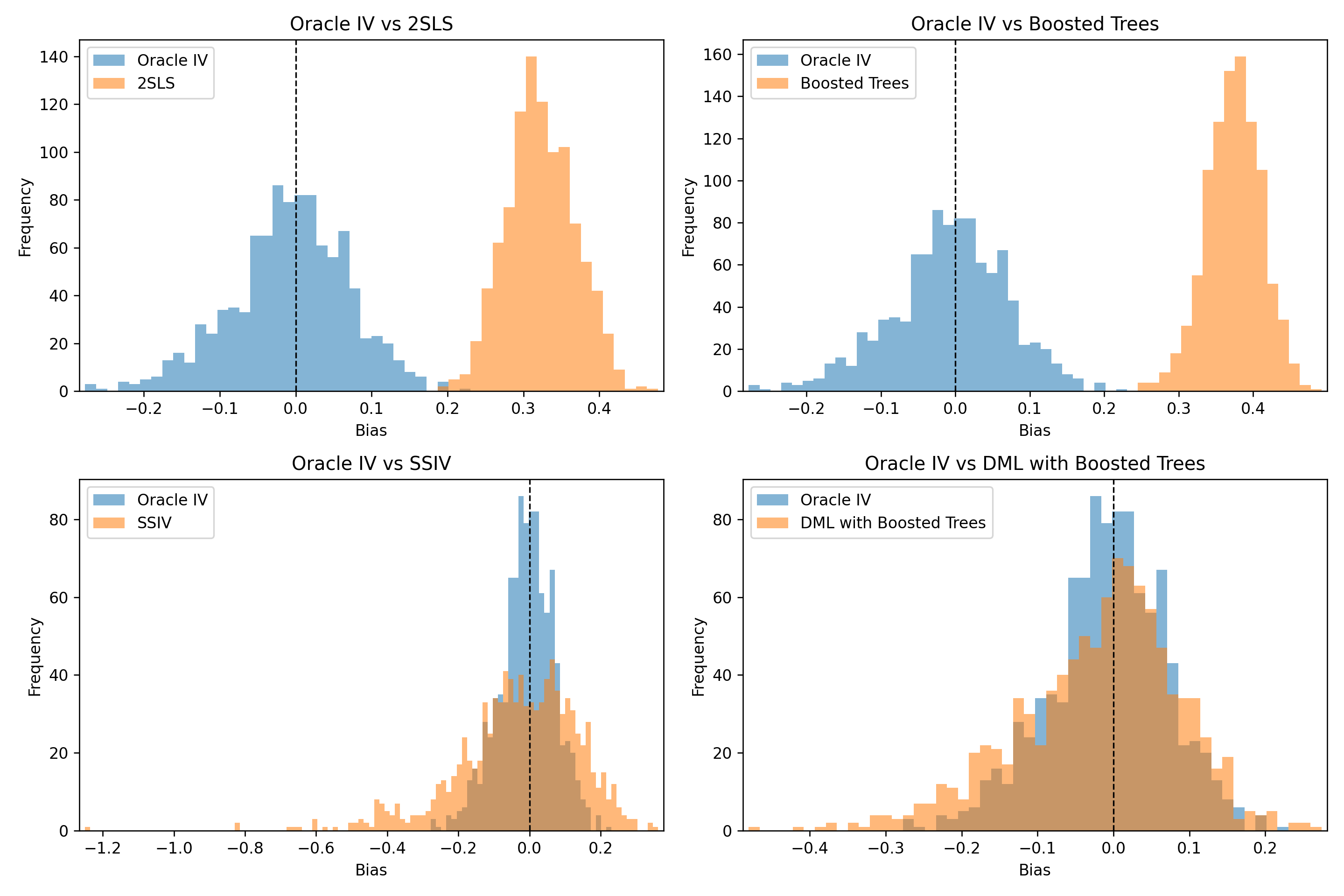}
 \parbox{\linewidth}{
 \scriptsize \textit{Notes.} Each panel compares simulation performance of a feasible IV estimator (orange) to the infeasible oracle IV estimator (blue). 
 }
\end{figure}

\begin{table}[!t]
    \footnotesize\singlespacing\centering
    \caption{IV simulation results by estimation method}\label{table:iv_sim_results}
    \begin{tabular}
    {lcccc}
    \toprule
    Variable & Bias & Median Bias & Std. Dev. & Coverage \\
    \midrule
    Oracle IV & -0.0076 & -0.0036 & 0.0773 & 0.9570 \\
    2SLS & 0.3228 & 0.3199 & 0.0447 & 0.0000 \\
    Boosted Trees & 0.3738 & 0.3751 & 0.0370 & 0.0000 \\
    SSIV & -0.0325 & -0.0172 & 0.1725 & 0.9380 \\
    DML with Boosted Trees & -0.0201 & -0.0051 & 0.1075 & 0.9590 \\
    \bottomrule
    \vspace{.1cm}
    \end{tabular}
    \parbox{\linewidth}{
    \scriptsize
    \textit{Notes:} This table presents simulation summary statistics of estimators $\theta_0$ as described in the main text. Coverage denotes the simulation coverage of 95\% confidence intervals based on homoskedastic standard errors. All estimators use Neyman orthogonal scores. SSIV and DML with Boosted Trees also use cross-fitting. 
    }
\end{table}

We report results in Figure~\ref{fig:iv_sim} and Table~\ref{table:iv_sim_results}. As expected and predicted by the theory, the methods that do not use cross-fitting---2SLS and Boosted Trees---perform poorly. Although the distributions of estimates are tightly concentrated, they are centered far from the truth, yielding average biases an order of magnitude larger than estimators relying on cross-fitting. As a result, their 95\% confidence intervals exhibit zero coverage across the 1000 simulation replications. In contrast, the estimators with cross-fitting, SSIV and DML with Boosted Trees, perform substantially better. They have relatively small bias and produce coverage near the nominal 95\% level. Compared to the infeasible oracle IV, they exhibit a visually larger spread, but are also approximately centered around zero.

Importantly, we observe a meaningful difference between the two cross-fitted estimators. DML with Boosted Trees outperforms SSIV, reflecting the advantage of using a more flexible and regularized first-stage learner in this context. While this is specific to the design considered here, it highlights a broader point: the choice of learner can matter substantially in practice. We return to this theme in the empirical example in Section~\ref{sec:application_monopsony}.

Finally, we emphasize that all five estimators rely on Neyman orthogonal scores. As such, these simulations are designed to isolate the impact of cross-fitting in reducing the impact of overfitting during nuisance parameter estimation. In the next subsection, we shift focus to the role of orthogonality itself by comparing estimators with and without Neyman orthogonal scores in the context of average treatment effect estimation.

\subsection{Average Treatment Effect Estimation}
\label{sec:example_401k}

%\thomas{Changed DR to AIPW to be consistent with Example~4. Double checking would be helpful.}

This section uses a calibrated simulation to illustrate the importance of Neyman orthogonality for valid inference. We focus on estimation of the average treatment effect (ATE) under unconfoundedness---that is, assuming treatment is as good as randomly assigned after conditioning on covariates. This setting is instructive as it allows comparison of DML to the IPW estimator, which is commonly used despite not being based on a Neyman orthogonal score.

The simulation is based on \citet{poterba1995}, who study the effect of 401(k) eligibility ($D$) on household net financial assets ($Y$), treating eligibility as random given observed covariates. The observed covariates ($X$), include age, income, education, family size, and indicators for two-earner households, home ownership, and alternative pension coverage. We use this application because it is a standard example in work on treatment effect estimation with machine learning; see, for instance, \citet{belloni2017program, chernozhukov2018, wuthrich2021, Ahrens2024_applied}.

%We base the simulation on the same 9915 observations used in the studies cited above. We estimate the probability of 401(k) eligibility given covariates, $r_0(X) = \Pr(D = 1 \mid X)$, and the conditional mean of financial assets given treatment and covariates, $\ell_0(D, X) = \Ep[Y \mid D, X]$, using random forests trained on the full sample. For this simulation exercise, we use 1000 trees and a minimum node size of 10 in each forest. We also estimate treatment-specific residual variances, $\sigma_d^2$, by computing the sample variance of $y_i - \hat\ell(d_i, x_i)$ within the treated and control groups, where $y_i$, $d_i$, and $x_i$ denote the observed outcome, treatment, and covariates for individual~$i$.

%In each simulation replication, we generate new treatment assignments and outcomes for all individuals in the original sample. The simulated treatment $d_{i,s}$ is drawn from a Bernoulli distribution with success probability $\hat r(x_i)$, where $x_i$ denotes the covariates. Given the simulated treatment, the outcome is generated as $y_{i,s} = \hat \ell(d_{i,s}, x_i) + \varepsilon_{i,s}$, with $\varepsilon_{i,s} \sim N(0, \hat \sigma^2_{d_{i,s}})$. Under this design, the true value of the ATE is approximately 6889.

We base the simulation on the same 9915 observations used in the studies cited above. We calibrate the data-generating process for the simulation by flexibly estimating the propensity score and the conditional outcome model. We then use the covariate values for each individual in the original data to simulate new treatment assignments and outcomes. Under this design, the true value of the ATE is approximately 6,889.\footnote{Specifically, we estimate the probability of 401(k) eligibility given covariates, $r_0(X)=\Pr(D=1\mid X)$, and the conditional mean of financial assets given treatment and covariates, $\ell_0(D,X)=\Ep[Y\mid D,X]$, using random forests trained on the full sample (1,000 trees and minimum node size 10). We estimate treatment-specific residual variances, $\sigma_d^2$, as the sample variance of $y_i-\hat\ell(d_i,x_i)$ within treatment groups. In each simulation replication, we draw $d_{i,s}\sim\text{Bernoulli}(\hat r(x_i))$ and generate outcomes as $y_{i,s}=\hat\ell(d_{i,s},x_i)+\varepsilon_{i,s}$, where $\varepsilon_{i,s}\sim N(0,\hat\sigma^2_{d_{i,s}})$.}

We compare three scores for estimating the ATE: the IPW score, the Neyman orthogonal AIPW score, and the regression adjustment (RA) score. The IPW and AIPW scores are defined in Example~4. The RA score is given by
\begin{align}\label{eq:late_reg_score}
    m_{RA}(W; \theta, \eta) = \ell(1,X) - \ell(0,X) - \theta,
\end{align}
where the nuisance parameter is $\eta(D,X) = \ell(D,X)$ with true value $\ell_0(D,X) = \Ep[Y\vert D, X].$ Like the IPW score, the RA score is not Neyman orthogonal. For each score, we report estimates both with and without cross-fitting. The DML estimator of the ATE corresponds to using the AIPW score with cross-fitting. All estimators use the same tuning parameter choices for nuisance function estimation, so differences in performance reflect only the choice of score and use of cross-fitting. We use random forests with 1000 trees and a maximum depth of 8 and 4 for the outcome and treatment propensity functions, respectively. We set the number of folds to $K=10$.

\begin{figure}[!t] 
\caption{Histograms of Cross-fit ATE Estimators }\label{fig:401k_sim}
    \centering
    \includegraphics[width=0.9\textwidth]{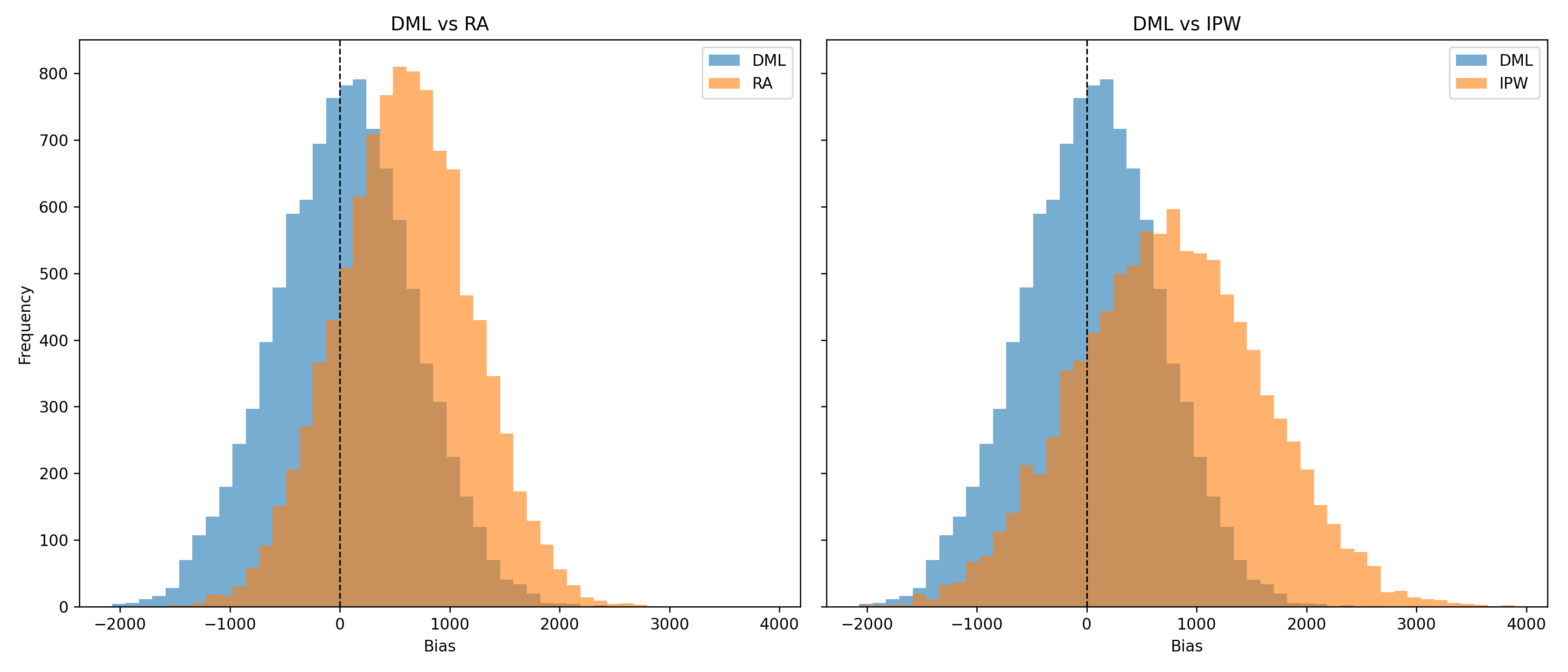}
 \parbox{\linewidth}{
 \scriptsize \textit{Notes.} Each panel compares simulation performance of the DML estimator (blue) to a cross-fit ATE estimator based on a score that is not Neyman orthogonal (orange). Specifically, the left panel compares DML to a cross-fit regression adjustment (RA) estimator, and the right panel compares DML to a cross-fit IPW estimator. 
 }
\end{figure}

Figure \ref{fig:401k_sim} is designed to highlight the role of Neyman orthogonality by comparing the DML estimator to two alternatives based on non-orthogonal scores: the cross-fit RA estimator (left panel) and the cross-fit IPW estimator (right panel). The most striking feature in both comparisons is that DML appears approximately unbiased, while the RA and IPW estimators exhibit substantial bias. The distribution of DML is also more concentrated than that of IPW and similar in spread to RA. Taken together, the plots show that DML dominates in this example due to its much smaller bias and comparable or better precision.

Table~\ref{tab:401k_res} reports simulation results across a range of performance metrics. The DML and AIPW estimators, which are each based on Neyman orthogonal scores, dominate the remaining procedures. They exhibit substantially lower bias and mean absolute deviation than the IPW and RA estimators, whether or not cross-fitting is used. While the RA estimators have a slightly smaller standard deviation than DML and AIPW, this is offset by their much larger bias, implying inferior performance for most reasonable loss functions.

DML and AIPW also exhibit superior coverage performance. For these two estimators, we report confidence intervals based on the standard error estimator defined in~\eqref{eq:Sigma_hat}. This estimator is theoretically justified for DML under relatively weak conditions and would be valid for AIPW if overfitting were sufficiently controlled. In contrast, for the non-orthogonal IPW and RA estimators, we report coverage using the simulation standard deviation, since their first-order behavior depends directly on the nuisance function estimator, which makes valid standard error estimation challenging.

Even using the infeasible simulation standard deviation, the IPW and RA estimators substantially undercover the true ATE. While substantially better than IPW or RA, we see that the AIPW estimator also undercovers. In contrast, DML achieves near-nominal coverage using estimated standard errors, and outperforms AIPW in both bias and coverage.\footnote{If we used simulation standard deviations for DML and AIPW, coverage would be 0.948 and 0.945.} This highlights the central insight of the simulation: valid inference relies on combining Neyman orthogonality with cross-fitting.

\begin{table}[!t]
\centering\footnotesize\singlespacing
\caption{ATE Simulation Results}
\label{tab:401k_res}
\begin{tabular}{lcccccc}
\toprule
 & $\hat\theta_{DML}$ & $\hat\theta^{\,\times \text{-fit}}_{RA}$ & $\hat\theta^{\,\times \text{-fit}}_{IPW}$ & $\hat\theta_{AIPW}$ & $\hat\theta_{RA}$ & $\hat\theta_{IPW}$ \\
\midrule
Bias & 47.3 & 594.9 & 756.4 & 142.0 & 644.0 & 747.3 \\
Median Bias & 60.8 & 597.8 & 755.1 & 154.7 & 647.1 & 745.6 \\
Mean Abs. Dev. & 502.2 & 701.1 & 923.7 & 508.8 & 734.4 & 914.9 \\
Std. Dev. & 627.4 & 606.2 & 841.7 & 619.2 & 604.7 & 837.3 \\
%Feasible Coverage & 0.9451 & 0.1998 & 0.9662 & 0.9110 & 0.2031 & 0.9654 \\
%Infeasible Coverage & 0.9484 & 0.8339 & 0.8513 & 0.9450 & 0.8152 & 0.8528 \\
Coverage & 0.945 & 0.834 & 0.851 & 0.911 & 0.815 & 0.853 \\
& & & & & & \\
Neyman orthogonal & Yes & No & No & Yes & No & No \\
Cross-fitting & Yes & Yes & Yes & No & No & No \\
\bottomrule
\end{tabular}
\par\medskip
\parbox{\linewidth}{\scriptsize{\it Notes:} $\hat\theta_{AIPW}$, $\hat\theta_{RA}$, and $\hat\theta_{IPW}$ respectively denote estimation based on the AIPW, RA, and IPW scores without cross-fitting. Similarly, $\hat\theta^{\,\times \text{-fit}}_{RA}$, and $\hat\theta^{\,\times \text{-fit}}_{IPW}$ respectively denote estimation based on the RA and IPW scores with cross-fitting. $\hat\theta$ is the DML estimator, which is based on the AIPW (Neyman orthogonal) score with cross-fitting. Coverage is simulation coverage of 95\% confidence intervals. For both $\hat\theta_{DML}$ and $\hat\theta_{AIPW}$, intervals are computed using the standard error estimator implied by \eqref{eq:Sigma_hat}. For the remaining estimators, intervals are computed using the infeasible simulation standard deviation. Results are based on 10000 simulation replications.
}
\end{table}

%Because the simulation is calibrated using simple random forests, applying the same learners for estimation should approximately satisfy DML's assumption of moderate-quality nuisance estimation. This setup helps ensure that differences in performance across estimators primarily reflect the choice of score and the use of cross-fitting, rather than variation in machine learning quality. In real applications, however, the choice of learner is both more difficult and potentially consequential. We return to this point in Section~\ref{sec:application_monopsony}.

%%%%%%%%%%%%%%%%%%%%%%%%%%%%%%%%%%%%%%%%%%%%%%%%%%%%%%%%%%%%%%%%%%%%%%%%%%%%%%%%%%%%%%%%%
%%%%%%%%%%%%%%%%%%%%%%%%%%%%%%%%%%%%%%%%%%%%%%%%%%%%%%%%%%%%%%%%%%%%%%%%%%%%%%%%%%%%%%%%%

\section{Economic Consequences of Hospital Admission}\label{sec:example_hrs}

To demonstrate the flexibility of DML, we apply it in a staggered adoption panel setting to estimate group-time average treatment effects on the treated and dynamic average treatment effects as discussed in \citet{callaway:santanna}. Outside of illustrating the application of DML in a canonical panel data setting, we further use this example to discuss the additional randomness introduced in DML due to the use of sample splitting. We show that simply repeating the DML estimation can help us gauge the robustness of conclusions to particular sample splits. We then outline the median aggregation approach presented in \citet{chernozhukov2018} as a way to summarize results across repetitions. 

Our example builds on \citet{dobkin2018economic} and \citet{sun2021estimating}. \citet{dobkin2018economic} analyze the causal effect of hospital admission on several economic outcomes, including out-of-pocket medical spending, using conventional two-way fixed effects applied to a panel of U.S. households from the Health and Retirement Study (HRS). \citet{sun2021estimating} extend this analysis by estimating dynamic effects using more flexible methods that allow for treatment effect heterogeneity. We use the same data as \citet{sun2021estimating}, which consist of a balanced panel of 656 households observed over waves~7 through 11 of the HRS.\footnote{See Table 2 of \citet{sun2021estimating} for summary statistics of the HRS sample we use.}

We proceed in two steps. In Section \ref{subsec:gt_att}, we estimate the change in out-of-pocket medical spending at time period $t$ caused by hospitalization at time $g$ for all potential pairs $(t, g)$ in the data. In Section~\ref{subsec:dyn_att}, we then aggregate these estimates for inference about the dynamic effects of hospitalization.

\subsection{Group-Time Average Treatment Effects of Hospitalization}\label{subsec:gt_att}

The group-time average treatment effect on the treated (GT-ATT) measures the effect of hospitalization at time $g$ on medical spending at time $t\leq T$, for individuals first hospitalized in period $g$. While the GT-ATTs are often not of primary interest, they can be aggregated to population-weighted ATTs or dynamic effects. We follow the potential outcome setup of \citet{callaway:santanna} to formally define the GT-ATT. Let $G_i$ denote the time of first hospital admission for individual $i$. Let $Y_{i,t}(0)$ denote the potential outcome of out-of-pocket medical spending of individual $i$ at time $t$ if they remain untreated throughout, and let $(Y_{i,t}(g))_{g=2}^T$ be the set of potential outcomes at time $t$ for every potential time $g$ of the first hospital admission. Observed outcomes are given by $Y_{i,t}=Y_{i,t}(0) + \sum_{g=2}^T (Y_{i,t}(g) - Y_{i,t}(0))\mathbbm{1}\{G_i=g\}$. Then, the GT-ATT for group $g$ and time $t$ is defined as \begin{align}
\begin{aligned}\label{eq:definition_ATTgt}
\theta_0^{%ATT
(g,t)} = \Ep[Y_{i,t}(g) - Y_{i,t}(0)\vert G_{i}=g].
\end{aligned}
\end{align}

Following \citet{dobkin2018economic} and \citet{sun2021estimating}, we leverage a parallel trends assumption on not-yet-hospitalized individuals to identify the GT-ATTs. Motivated by the robustness checks of \citet{dobkin2018economic}, we consider a more flexible version of this assumption: parallel trends are assumed to hold only after conditioning on observed pre-treatment characteristics, such as age, gender, race, and education. Under this conditional parallel trends assumption and additional standard conditions \citep[see, e.g.,][]{callaway:santanna}, the GT-ATT is equivalent to \begin{align*}
\begin{aligned}
\theta_0^{%ATT
(g,t)} &=\Ep[\Delta_g Y_{i,t}\vert G_i=g]  - \Ep[\Ep[\Delta_g Y_{i,t}\vert G_i\ne g, G_i>t,  X_{i}]\vert G_i=g],
\end{aligned}
\end{align*}
where $\Delta_g Y_{i,t} = Y_{i,t} - Y_{i,g-1}$ is the difference of outcomes in period $t$ and the baseline period $g-1$. A standard no anticipation assumption implies $\theta_0^{
(g,t)} = 0$ for $t < g$, which is often used to test for pre-trends. 

A key insight about the estimation of the GT-ATT is that for each group-time pair, the target parameter is equivalent to a conventional ATT identified under conditional unconfoundedness where the outcome is replaced by an appropriate difference of outcomes. Building on the efficient ATT score of \citet{hahn-pp}, %\citet{chang2020double} proposes a DML estimator based on the AIPW score for the GT-ATT:\footnote{See also \citet{santanna2020}.} 
we employ an augmented inverse probability weighted (AIPW) score for the group–time average treatment effect (GT-ATT) as the basis for DML:\footnote{Our treatment mirrors \citet{callaway:santanna}. The AIPW score directly builds on \citet{chang2020double}, who proposes a DML estimator in the canonical two-group, two-period difference-in-differences design. See also \citet{santanna2020}.}
\begin{align}\label{eq:DR_GTATT}
\begin{aligned}  
    m^{(g,t)}(W_i; \theta, \eta) 
    &= \frac{\mathbbm{1}\{G_i=g\}\big(\Delta_g Y_{i,t} - \ell^{(g,t)}(X_i)\big)}{\pi^{g}} \\
    &\quad - \frac{q^{(g,t)}(X_{i})\,\mathbbm{1}\{G_i \ne g\}\mathbbm{1}\{G_i>t\}\big(\Delta_g Y_{i,t} - \ell^{(g,t)}(X_i)\big)}{\pi^{g}\big(1-q^{(g,t)}(X_i)\big)} \\
    &\quad - \frac{\mathbbm{1}\{G_i=g\}}{\pi^{g}}\,\theta
    \end{aligned}
\end{align}
where $\mathbbm{1}\{G_i=g\}$ denotes the treated group, $\mathbbm{1}\{G_i \ne g\}\mathbbm{1}\{G_i>t\}$ denotes units not yet treated by time $t$ that serve as controls, and the nuisance parameter $\eta = \big(q^{(g,t)}, \ell^{(g,t)}, \pi^{g}\big)$ takes true values 
$q_0^{(g,t)}(X_i) = \Pr\big(G_i=g \mid X_i,\{G_i = g\} \cup \{G_i > t\}\big),$ 
$\ell_0^{(g,t)}(X_i)= \Ep[\Delta_g Y_{i,t}\mid G_i \ne g,\ G_i>t,\ X_i],$ 
and $\pi^{g}_0 = \Pr(G_i=g)$.
One can confirm that the score \eqref{eq:DR_GTATT} satisfies both the moment condition \eqref{eq: score_identification} and Neyman orthogonality (using steps similar to Example 4, Appendix \ref{app: neyman verification}).

The application of DML to estimation of the GT-ATTs provides a useful robustness check because it seems unlikely that a researcher has prior knowledge of the parametric form of the nuisance functions. Even in this application with only few control variables, it is practically impossible to saturate the model, which would eliminate any additional need for flexible estimation. A fully interacted set of the eight control variables results in 3,072 indicators, which exceeds the number of households in the sample. 

We set the number of cross-fitting folds to $K = 15$. Since the number of observations per group-time sample is relatively small, we opt for more folds than the rule of thumb of 5-10 folds often cited in the literature on cross-validation. For ease of exposition, we report only results on a single machine learner: a random forest estimator with 1000 trees and a minimum node size of 10 to estimate the nuisance parameters.

Table~\ref{tab:hrs_attgt} presents DML estimates for all identified group-time average treatment effects. Columns~(1) through (5) show results from five different random partitions of the data used for cross-fitting. Although these DML estimators are asymptotically equivalent, they can yield different results in finite samples due to variation across sample splits. This variability may be particularly pronounced in settings with relatively few observations, as is often the case in staggered adoption designs. 

To illustrate, consider the first-period treatment effect for group 9. This estimate is based on only 176 treated and 228 untreated observations. The corresponding DML estimates obtained from the five different considered sample splits range from 3204 to 3486---a difference of roughly 30\% of the corresponding standard errors. While this variation is not large enough to change the sign or significance of the estimate, it shows that DML can produce meaningful differences in practice depending on the sample split.

We believe that replicating the DML procedure several times is important for understanding the variability induced by sample splitting. While under ideal conditions the DML estimator from any one split is asymptotically equivalent to that from any other, finite-sample differences may arise. More importantly, large variation across splits may signal deeper issues. For example, substantial across-split variation may signal that asymptotic approximations are unreliable in the sample at hand (e.g., due to heavy tails, poor overlap, or unstable nuisance estimates), or that the underlying machine learners are unstable within the available data.

However, reporting all results may lead to information overload or be practically infeasible. We therefore recommend that researchers examine the full set of estimates and standard errors across splits and ensure they are available for transparency. Following \citet{chernozhukov2018}, we suggest median aggregation as a simple and practical strategy for summarizing results.\footnote{\citet{chernozhukov2018} also consider mean aggregation. We are not aware of formal arguments favoring one approach over another in finite samples. Our recommendation is heuristic.} Letting $\hat{\theta}_s$ and $\widehat{s.e.}_s$ denote the $s^{\textnormal{th}}$ DML estimator and its standard error across $S$ replications, the median-aggregated point estimate and standard error are
\begin{align}
\begin{split}\label{eq:median_aggregation}
    \hat{\theta}^{\text{median}} &= \text{median}\left(\{\hat{\theta}_s\}_{s=1}^S\right), \\  \widehat{s.e.}^{\text{median}} &= \sqrt{\text{median}\left(\{\widehat{s.e.}_s^2 + (\hat{\theta}_s-\hat{\theta}^{\text{median}})^2\}_{s=1}^{S}\right)},
\end{split}
\end{align}
where the median is applied element-wise for vector-valued parameters. The standard error accounts for both conventional sampling uncertainty and variability across splits.

Column (6) of Table \ref{tab:hrs_attgt} presents the median-aggregated estimators for the identified GT-ATTs. The results strengthen the economic conclusions of \citet{dobkin2018economic} and \citet{sun2021estimating}. When we use DML to estimate the same effects allowing for parallel trends to hold only conditional on controls and using a flexible method to include those controls, we continue to estimate that hospitalization causes a substantial and significant increase in out-of-pocket medical spending in the period immediately following the event. This conclusion holds for all three cohorts considered in our analysis.

We also compare the DML estimates to those obtained using the parametric AIPW estimator proposed by \citet{santanna2020}. This estimator relies on the same AIPW score as DML but differs in how the nuisance functions are estimated. Specifically, it assumes that the propensity score $\Pr\big(G_i=g \mid X_i,\{G_i = g\} \cup \{G_i > t\}\big)$ follows a logit model with a linear index and that the outcome regression $\Ep[\Delta_g Y_{i,t}\mid G_i \ne g,\ G_i>t,\ X_i]$ is linear in $X_i$. If both parametric models are correctly specified, the estimator is first-order equivalent to DML. If either is misspecified, inference remains valid but efficiency may be lost. If both are misspecified, the estimator is inconsistent. Because these parametric forms are typically motivated by convenience rather than theory, DML offers an appealing alternative. In practice, the choice between the two depends on the researcher’s confidence in the parametric models and access to a flexible nuisance estimator of reasonable quality. Regardless of the preferred approach, using the other as a robustness check is a sensible practice.

Columns (7) and (8) of Table \ref{tab:hrs_attgt} present the \citet{santanna2020} estimator without and with controls, respectively. The comparison confirms that adjusting for covariates materially affects the results. Relative to the size of the standard errors, the median-aggregated DML estimates (in Column~6) and the AIPW DiD estimates (Column~8) are qualitatively similar. The most notable exception is group 9's treatment effect in Wave 10. The parametric estimate is 2534 ($s.e.=422$) and thus substantially larger than the DML estimate of 937 ($s.e.=434$). Further, group 10's pre-treatment effect in Wave 7 is negative under DML ($-1492$, $s.e.=2223$) but positive under the parametric estimator ($1222$, $s.e.=1127$), though both are imprecisely estimated due to the small cohort size. While the true effects are unknown, the discrepancy highlights the potential for results to strongly differ based on nuisance function specification. We return to this issue in Section~\ref{sec:application_monopsony}.

\begin{table}[!t]
    \scriptsize\singlespacing\centering
    \caption{Group-Time Average Treatment Effects on the Treated by Estimator}\label{tab:hrs_attgt}
\input{tables/hrs/attgt_ranger2}\par\medskip
    \parbox{\linewidth}{
      \scriptsize
      \textit{Notes:} Columns (1)-(5) show DML GT-ATT estimators for randomly generated cross-fitting folds. Column (6) presents the corresponding median aggregated DML estimate. Columns (7) and (8) present the \citet{santanna2020} estimator without and with controls, respectively. Point-wise standard errors are in parentheses. 
      }
\end{table}

\subsection{Dynamic Effects of Hospitalization}\label{subsec:dyn_att}
In many applications, researchers are primarily interested in summaries of group-time effects. A leading example is the dynamic treatment effect, which underpins many event study designs and provides a way to assess the plausibility of parallel trends.

Formally, dynamic effects are group-weighted averages of the GT-ATTs:
\begin{align}\label{eq:definition_DATTt}
\tau_0^{(e)} = \sum_{g \in \mathcal{G}} \mathbbm{1}\{g + e \leq T\} \Pr(G_i = g \mid G_i + e \leq T) \theta_0^{%ATT
(g, g+e)},
\end{align}
where $\mathcal{G}$ is the set of all treatment initiation periods, and $\theta_0^{(g,t)}$ is the group-time ATT defined earlier. In our context, this corresponds to averaging the effect of hospitalization across all groups observed $e$ periods after their initial admission, weighted by group size. That is, the dynamic effect $\tau_0^{(e)}$ measures the average effect of hospitalization $e$ periods after the initial admission.

Estimating the probabilities $\Pr\left(G_i=g \vert G_i + e \leq T\right)$ in \eqref{eq:definition_DATTt} is straightforward. For example, a natural estimator is given by the binning estimator $\sum_{i=1} \mathbbm{1}\{G_i=g, G_i + e \leq T\}/\sum_{i=1} \mathbbm{1}\{G_i + e \leq T\}$. Given an estimator for these probabilities and the DML estimators for the GT-ATTs, inference for $\tau_0^{(e)}$ then follows from standard results about inference on linear combinations.

\begin{figure}[!t] 
\caption{Dynamic Average Treatment Effect Estimates}\label{fig:hrs_dyn}
    \centering\singlespacing
    \includegraphics[width=0.7\textwidth]{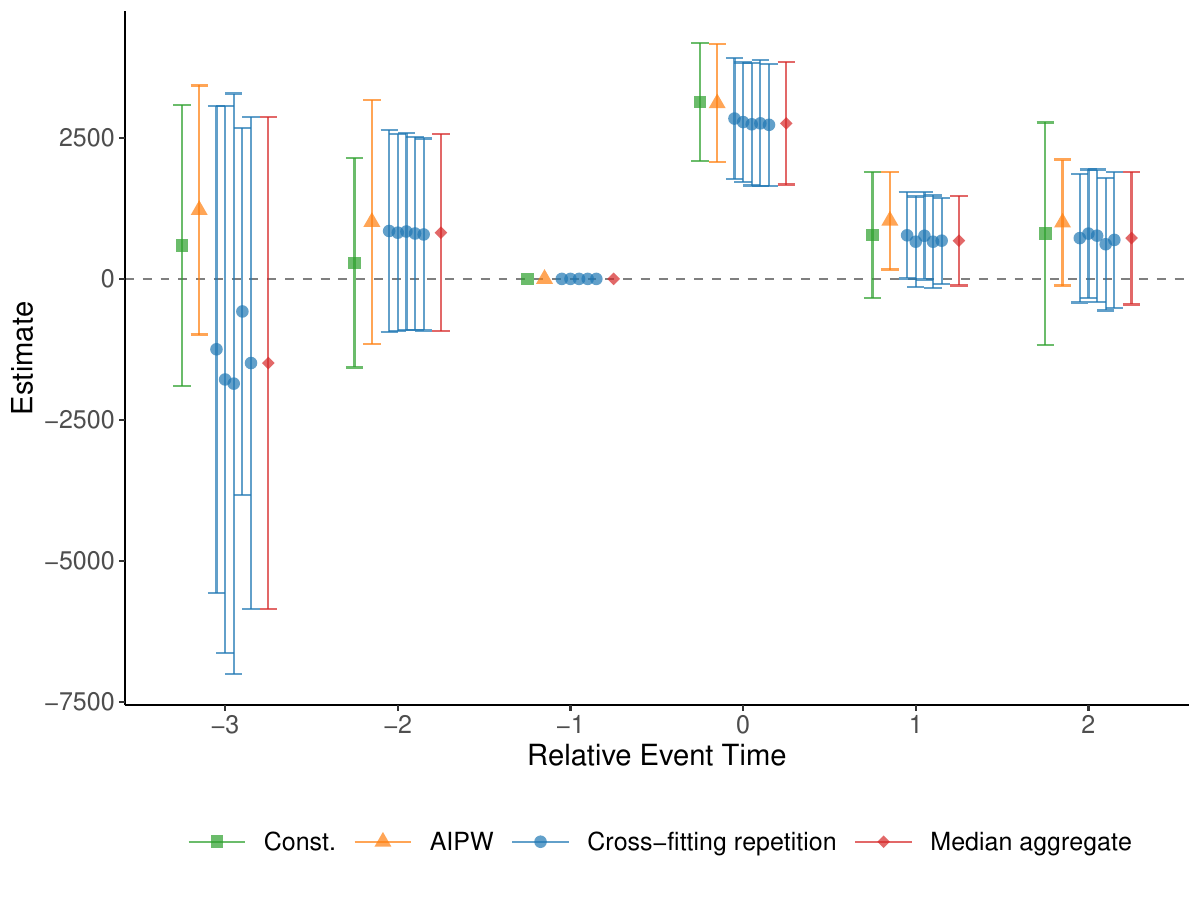}
     \vskip0.25em
     \parbox{0.9\textwidth}{\scriptsize \textit{Notes.} The figure displays dynamic treatment effect estimates using different GT-ATT estimators. ``Const.'' and ``AIPW'' refer to the \citet{santanna2020} estimator without and with controls, respectively. ``Cross-fitting repetition'' shows DML GT-ATT estimates across random cross-fitting folds, and ``Median aggregate'' presents the corresponding median aggregated DML estimate. Bars indicate point-wise 95\% confidence intervals.}  
\end{figure} 

Figure \ref{fig:hrs_dyn} displays the estimated dynamic effects of hospitalization obtained from the different GT-ATT estimators considered previously. Across methods, we find no statistically significant effects in the pre-treatment period and a large, significant increase in out-of-pocket medical spending immediately following hospitalization. DML estimates and the parametric AIPW alternative are quantitatively similar overall, and the variation in DML across different cross-fitting splits is quite small.

One notable difference is at event time $-3$: the DML estimate is substantially less precise than the parametric alternative. Because only group 10 has outcomes observed three periods before hospitalization, this dynamic effect corresponds exactly to group 10's pre-treatment effect in Wave 7, which we discussed in the preceding section. Since this estimate is based solely on 163 treated and 65 never-treated observations, it is not surprising that it is relatively imprecise. That it is markedly less precise than the corresponding parametric estimate highlights how, with limited data, functional form restrictions can meaningfully shape inference. The discrepancy illustrates how DML can serve as a robustness check on parametric inference when data are limited. At a minimum, it cautions against taking the parametric results at face value without careful economic justification for the functional form assumptions.

We report results for other learners and tuning parameter choices in the companion website for this paper, and find similar results. For example, using random forests with higher regularization (minimum node size of 100) yields a first-period treatment effect estimate of 2903 ($s.e.=497$), while lower regularization (minimum node size of 1) yields 2666 ($s.e.=571$). Pre-treatment effects remain statistically insignificant at the 5\% level across all learner and tuning parameter choices considered. 

The robustness to learner choice is quite different in the example we consider in next section. There we revisit the study on monopsony power of \citet{dube2020} and illustrate that the selection and tuning of nuisance estimators can be empirically consequential. We use this application to demonstrate diagnostic analyses that can assist in evaluating and selecting nuisance estimators.

%%%%%%%%%%%%%%%%%%%%%%%%%%%%%%%%%%%%%%%%%%%%%%%%%%%%%%%%%%%%%%%%%%%%%%%%%%%%%%%%%%%%%%%%%
%%%%%%%%%%%%%%%%%%%%%%%%%%%%%%%%%%%%%%%%%%%%%%%%%%%%%%%%%%%%%%%%%%%%%%%%%%%%%%%%%%%%%%%%%

\section{Monopsony in Online Markets}\label{sec:application_monopsony}

We revisit the \citet{dube2020} paper on monopsony power on Amazon MTurk, an online platform for hiring workers to perform tasks. The research falls into the wider literature estimating the labor supply elasticity \citep[see reviews in][]{Sokolova2021,langella2021}. This example is useful for at least two reasons. First, it illustrates how DML can incorporate complex, non-tabular data, such as text or images, which are increasingly common in empirical economics \citep[e.g.,][]{ash2023a}. Second, it provides an interesting setting for discussing the choice of ML method for nuisance estimation. In particular, it highlights the importance of selecting learners and tuning parameters to support credible inference. We use this example to discuss methods for evaluating such choices.

\citet{dube2020} consider multiple identification strategies and datasets for studying monopsony power. We focus here on their analysis of a cross-sectional sample compiled by \citet{ipeirotis2010analyzing}. The parameter of interest is the partially linear regression coefficient $\theta_0$ introduced in Example~2. The outcome variable, $Y$, is the logarithm of the time required for a posted task to be filled. The treatment variable, $D$, is the logarithm of the payment offered. The covariates, $X$, are a high-dimensional vector of task characteristics.

The authors provide arguments under which $\theta_0$ is the negative labor supply elasticity to the firm and thus a measure of monopsony power. Since tasks vary in complexity, difficulty, and time commitment, it is important to adjust for characteristics that are likely correlated with duration and reward. Some of these task characteristics---such as title, task description, and keywords---come in the form of text. To account for these, the authors include a combination of hand-engineered features and vectorized representations of the task's title, description, and keywords. The hand-engineered features include, among others, the allotted time, common patterns identified using regular expressions, the number of keywords, and the lengths of the title and description. The vectorized text representations used by \citet{dube2020} consist of Doc2Vec embeddings, topic distributions, and n-grams.

In our reanalysis, we retain the original hand-engineered covariates but replace the text representations used by \citet{dube2020} with fine-tuned embeddings from the large language model DeBERTa v3. Large language models map text into numerical representations, referred to as embeddings, that capture semantic and contextual information. We use fine-tuning to adapt the pre-trained parameters to our setting, thereby improving predictive performance.\footnote{We implement fine-tuning using Low-Rank Adaptation \citep[LoRA; ][]{hu2021loralowrankadaptationlarge}. Rather than fully retraining all parameters, LoRA optimizes a low-rank decomposition of the parameters updates, effectively modifying the model within a regularized subspace.} Fine-tuning is performed separately for the outcome and treatment models in each cross-fitting iteration, alleviating concerns about inducing over-fitting bias.

We consider 12 candidate learners that are trained on hand-coded controls and fine-tuned embeddings: OLS as a simple unregularized baseline, cross-validated lasso and ridge as regularized linear methods, and, to allow for nonlinearities, three implementations of random forests, three implementations of gradient boosting (\texttt{XGBoost}), and three feed-forward neural network architectures. The tuning parameters are reported in Table~\ref{tab:dube_monopoly_individual_learners}. We set the number of folds to $K=3$, repeat cross-fitting $S=5$ times, and report the median aggregate estimates. Finally, to account for possible dependence in the data, we implement cross-fitting by recruiter, assigning all observations from a given recruiter to the same fold, and compute standard errors clustered at the recruiter level.\footnote{The Online Appendix provides additional implementation details, results for $K = 5$, and results obtained when ignoring dependence in the construction of cross-fitting folds.}

    \begin{table}[!t]
    \centering
    \caption{Estimation results for the Monopsony application: Individual candidate learners}
    \label{tab:dube_monopoly_individual_learners}
    \scriptsize\singlespacing
    \begin{tabular}{l c c c c c c}
    \toprule\toprule
    & \multicolumn{6}{c}{\multirow{2}{*}{{\it Dependent variable:} log duration}} \\
     & \\
     {\it Panel A.}  & (1) & (2) & (3) & (4) & (5) & (6)\\
    \midrule
    \partialinput{10}{19}{dube_monopsony/output_xclust_cluster-robust_part1_K3} \\
    {\it Panel B.}  & (7) & (8) & (9) & (10) & (11) & (12) \\
    \midrule
    \partialinput{10}{19}{dube_monopsony/output_xclust_cluster-robust_part2_K3}\\[-.3cm]
    \bottomrule\bottomrule
    \end{tabular}
    \par\medskip
    \parbox{\linewidth}{{\it Notes:} 
    The table reports DML estimates based on 12 different nuisance estimators. The controls are the hand-coded controls of \citet{dube2020} and DeBERTa embeddings, fine-tuned using LoRA. We use $K=3$ cross-fitting folds, implemented by randomly assigning recruiters to folds, and report median aggregated estimates obtained using $S=5$ cross-fitting repetitions. The number of observations is $N=258,352$.
    The diagnostics reported are the cross-fitted $R^2$, the CVC $p$-value, and the short-stacking weights; each averaged across cross-fitting repetitions. We consider the following nuisance estimators: OLS; CV-lasso (lasso with tuning parameter selected by cross-validation); CV-ridge (ridge with tuning parameter selected by cross-validation); three types of random forest labeled RF~1 (600 trees), RF~2 (600 trees, minimum terminal node size of 500) and RF~3 (600 trees, minimum terminal node size of 2000); three types of \texttt{XGBoost} labeled XGB~1 (800 trees, early stopping after 10 rounds), XGB~2 (800 trees, minimum node size of 500) and XGB~3 (800 trees, minimum node size of 2000). DML estimation uses the \textsf{R} package \texttt{ddml} \citep{ddml_R}. The nuisance estimators were implemented with \texttt{glmnet} \citep{glmnet}, \texttt{ranger} \citep{ranger}, \texttt{XGBoost} \citep{xgboost}, and \texttt{keras} \citep{keras3R}. Standard errors are clustered at the recruiter level. 
    }
    \end{table}

Table~\ref{tab:dube_monopoly_individual_learners} shows that DML estimates can be highly sensitive to the choice of nuisance estimator, with point estimates ranging from $-3.9$ to $2.1$. These differences highlight a practical challenge with DML: ML offers a rich set of methods, but different learners may yield qualitatively distinct results. This challenge has been noted in the literature. For example, through calibrated simulations, \citet{Ahrens2024_applied} illustrate stark differences in inferential results for target parameters estimated with DML with different nuisance estimators; see also \citet{wuthrich2021}, \citet{giannone2021economic}, \citet{AF:MachineLabor}, and \citet{bach2024} for related discussions.

A convenient byproduct of cross-fitting is evidence about the out-of-sample performance of nuisance estimators. To provide guidance on the choice of nuisance estimator, we first consider the estimated out-of-sample $R^2$ values calculated using the cross-fitted predicted values and provided in the same table. These values indicate substantial variation across learner specifications.\texttt{XGBoost} (columns 7-9) achieves $R^2$ scores of around $85$\% and $74$\% for predicting outcome and treatment, respectively, whereas other learners perform markedly worse. The \texttt{XGBoost}-based estimates imply a labor supply elasticity between $0.040$ and $0.061$ ($s.e.\approx 0.022$). By contrast, the neural networks achieve outcome $R^2$ values below $4$\% and treatment $R^2$ values below $22$\%, and their corresponding point estimates display considerable instability.

The out-of-sample $R^2$ values provide a useful diagnostic but do not directly indicate whether differences across learners are statistically meaningful. To supplement this, Table~\ref{tab:dube_monopoly_individual_learners} also reports $p$-values from the cross-validation with confidence test \citep[CVC;][]{lei2020}. The null hypothesis in the CVC test is that a given learner achieves the lowest predictive risk among the full set of candidates. The alternative is that at least one other learner has a lower risk. \citet{lei2020} recommends forming a confidence set of the best learners by retaining those for which the $p$‑value exceeds a pre‑specified threshold, typically 0.1 or 0.2. Using a threshold of 0.2, the CVC tests identify learner specifications 8 and 9 (\texttt{XGBoost}~2 and~3) as the only candidate learners for the outcome equation and treatment equation for which the null cannot be rejected.

Rather than selecting the best-performing nuisance estimators based on the $R^2$ score or the CVC test, we can also combine them through stacking or model averaging approaches \citep{wolpert1996,Breiman1996a,vanderlaan:super}. Stacking constructs a ``super learner'' as a weighted average of candidate learners, where the weights are chosen to minimize out-of-sample prediction error. Within the DML framework, the stacking weights can be re-estimated at each step of the cross-fitting procedure, or estimated once using the full sample by regressing the outcome on the learners' cross-fitted predicted values (see \citealt{Ahrens2024_applied} for a discussion of various options).

In this application, we combine all 12 learners using a method commonly employed in stacking applications, namely constrained least squares: We regress outcome and treatment against cross-fitted predicted values while imposing that the coefficients on the learner predicted values are non-negative and sum to one, and use these estimated coefficients as weights to construct ``super learners'' for the outcome and treatment. We opt to estimate the stacking regression only once for outcome and treatment on the full sample. The weights are reported in Table~\ref{tab:dube_monopoly_individual_learners}. The model averaging procedure assigns large weights only to the three \texttt{XGBoost} specifications. The procedure produces a DML estimate of $-0.054$ with $s.e.=0.020$; see Table~\ref{tab:dube_monopoly_meta}. We obtain similar point estimates when selecting the single best learner for each nuisance function or when assigning equal weights to all learners selected by the CVC test. Our preferred estimates are thus consistent with \citet{dube2020}: using several samples, their DML estimates suggest a labor supply elasticity in the 0.0299---0.198 range.

    \begin{table}[htb]
    \centering
    \caption{Estimation results for the Monopsony application: Meta learning approaches}
    \label{tab:dube_monopoly_meta}
    \scriptsize\singlespacing
    % Note: Dube et al. point estimate is taken from paper, Table 1
    % R-squared are from `residuals_full_ipeirotis.dta` in the replication data, see Stata variable labels
    \begin{tabular}{l c c c}
    \toprule\toprule
     & \multicolumn{3}{c}{\multirow{2}{*}{{\it Dependent variable:} log duration}} \\
     & \\
      & Stacking & Single-best  & CVC \\
    \midrule
    \partialinput{10}{14}{dube_monopsony/output_xclust_cluster-robust_part3_K3}\\[-.3cm]
    \bottomrule\bottomrule
    \end{tabular}\par\medskip
    \parbox{\linewidth}{{\it Notes:} 
    The table reports DML estimates when several learners are combined by constrained least squares (imposing non-negative weights that sum to one), by selecting the single best learner per equation, or by assigning equal weights to all learners for which the CVC $p$-values are larger than 0.2. We use $K=3$ cross-fitting folds, implemented by randomly assigning recruiters to folds, and report median aggregated estimates obtained using $S=5$ cross-fitting repetitions. Standard errors are clustered at the recruiter level. 
    }
    \end{table}

The results in this application highlight that DML estimates can be sensitive to the choice of nuisance estimator. This sensitivity is consistent with conditions in existing DML theory, which indicate that poorly tuned or ill-suited learners can yield misleading results. We emphasize that the choice of which learner to use for a particular nuisance function and data set is rarely, if ever, known \textit{ex ante}. Further, there is no reason to assume that the same learner is best for every nuisance function, although this is implicitly imposed by many common estimation strategies. In practice, we recommend using a diverse set of nuisance function estimators to increase the credibility of DML estimates. Appropriate tools for evaluating these choices include cross‑fitted performance metrics such as $R^2$, the CVC test, and model averaging approaches.

%%%%%%%%%%%%%%%%%%%%%%%%%%%%%%%%%%%%%%%%%%%%%%%%%%%%%%%%%%%%%%%%%%%%%%%%%%%%%%%%%%%%%%%%%
%%%%%%%%%%%%%%%%%%%%%%%%%%%%%%%%%%%%%%%%%%%%%%%%%%%%%%%%%%%%%%%%%%%%%%%%%%%%%%%%%%%%%%%%%
\section{Discussion}\label{sec:discussion}

DML provides a flexible framework for inference in the presence of high-dimensional nuisance parameters. By combining Neyman orthogonal scores with cross-fitting, it mitigates the impact of estimating nuisance parameters, enabling asymptotically valid inference under relatively weak conditions. These conditions are compatible with a wide range of machine learning methods, making DML particularly useful in applications involving complex non-tabular data such as text or images. More broadly, the ability to leverage flexible estimators makes DML attractive as a complementary robustness check or when researchers wish to avoid parametric assumptions imposed for convenience rather than grounded in economic reasoning.

While DML provides robustness to the estimation of nuisance parameters, it is not a panacea. It does not tell the researcher what interesting target parameters are or how to identify them. Rather, it complements careful reasoning about objects of interest and their identification.

\paragraph{Implementation Guidance.}
When implementing DML, researchers face several design choices. One key decision is how to partition the data. With independent observations, forming folds at random provides a natural default. More generally, researchers should aim to form approximately independent folds by using partitions that respect any underlying dependence structure in the data, as noted in Remark~\ref{rem: dependence}. 

Researchers also need to select the number of folds $K$. Standard asymptotic results apply for any fixed $K$; see, e.g., \citet{chernozhukov2018}. These results suggest the use of relatively small values for $K$, but do not provide more specific guidance. \citet{velez:DML} provides a higher-order analysis within a restricted class of nuisance estimators and finds that performance improves with more folds, peaking at $K = n$ but with diminishing returns. Simulation evidence, including that in Section~\ref{sec:simulations}, suggests that conventional choices like $K = 5$ or $10$ work well in many settings. Given this evidence and that computational complexity is often a concern in applications of DML, we recommend choosing a simple round number informed by available computational resources. When the sample size is very small, larger values of $K$ may be desirable, and if resources permit, sensitivity analysis to $K$ is worthwhile.

To address the algorithmic randomness introduced by sample-splitting, we recommend simply repeating the cross-fitting procedure multiple times and reporting summaries of the resulting estimates, as illustrated in Section~\ref{sec:example_hrs}. This simple procedure reduces unappealing dependence on a particular random split and also serves as a useful diagnostic. Large variation across repetitions may raise concerns about finite-sample behavior or the plausibility of underlying assumptions, while stability across repetitions provides reassurance that results are not sensitive to a particular random split. 

Choice of nuisance function estimator is a first-order concern. In Section~\ref{sec:application_monopsony}, we showed that results can vary substantially across learners. %For example, if we had restricted our attention to regularized regression, our key result would have been an estimate of the labor supply elasticity more than five times larger than the DML estimates obtained using a learner selected after performing systematic robustness checks. 
Because it is rarely clear \emph{ex ante} which learner will perform best in a given application, we recommend considering a diverse set of candidate methods, including parametric benchmarks (e.g., linear or logistic regression), regularized regression, tree-based models, and neural networks. From a practical perspective, greater confidence in DML results is warranted when different learners deliver similar estimates, while substantial divergence across learners should be viewed as a warning sign that calls for caution and further investigation. We illustrated how predictive diagnostics, which are a natural byproduct of cross-fitting, can help identify potentially problematic learners. We encourage researchers to report their candidate learners, tuning procedures, and diagnostic metrics to promote transparency and replicability.

%Finally, we recommend empirically calibrating nuisance function estimates, as in \citet{laanDoublyRobustInference2025}, and, more broadly, using known functional restrictions. The approach in \citet{laanDoublyRobustInference2025} adapts the concept of calibration from predictive modeling, requiring that estimated nuisance functions satisfy known properties. As an example, if the nuisance function is $\eta_0(X) = \Ep[Y|X]$, we know that we should not be able to improve mean squared prediction error by taking transformations of $\eta_0(X)$. \citet{laanDoublyRobustInference2025} propose methods to enforce such calibration empirically and show that, for a broad class of target parameters, DML estimators remain asymptotically normal even if only one, as opposed to all, nuisance function estimator satisfies the usual convergence rate requirement. More generally, imposing known functional restrictions may substantially reduce the complexity of nuisance function estimation which should improve the performance of DML. It may be interesting to pursue further work in this direction.

\paragraph{Challenges and Caveats}
While DML is conceptually straightforward, its practical implementation can be challenging, and results can be highly sensitive to implementation choices. Decisions such as how nuisance functions are estimated, how many folds are used, how often cross-fitting is repeated, and how tuning parameters are selected can materially affect empirical results. Developing a deeper understanding of the finite-sample behavior of DML---including tradeoffs in learner complexity, computational complexity, fold count, and cross-fitting repetition---would be valuable not just for refining our theoretical knowledge but for guiding applied practice. 

From a theoretical standpoint, part of DML's appeal is that it delivers asymptotically normal inference even when flexible methods are used to estimate nuisance functions. Crucially, provided the nuisance estimator converges sufficiently quickly, the limiting distribution does not depend on the specific choice of nuisance estimator. Of course, it is unrealistic to expect all nuisance estimators to yield sufficiently accurate estimates in all settings. Many modern algorithms do not have theory establishing sufficiently fast rates without strong assumptions. For example, the results in \citet{CVFL:hdrandomforests} suggest that random forests, when applied to high-dimensional data with tuning similar to common defaults, may fail to converge quickly enough for DML, and results in \citet{cattaneo:trees} establish that deep regression trees may fail to be pointwise consistent. Further work establishing combinations of data-generating processes, learners, and tuning strategies that provide rates of convergence compatible with the sufficient conditions for DML is important for clarifying when, where, and how different learners should be used.

Convergence rate conditions may also fail in settings where nuisance functions are highly complex or non-smooth. Extending DML to better handle such settings is an important direction in current research; see, for example, \citet{robins2017minimax}, \citet{SPwsparsity}, and \citet{zheng2025perturbedDML}. 
Relatedly, the literature on sensitivity analysis in DML settings \citep[e.g.,][]{longstory} develops tools that, while motivated by concerns about unobserved confounding, are applicable more broadly. These tools can readily be adapted to assess robustness to other forms of misspecification, including insufficient learner flexibility or excessive nuisance function complexity. 

Finally, while accommodating dependent data is conceptually straightforward within the DML framework, its practical implementation raises additional challenges. Forming appropriate partitions, assessing effective sample sizes, and evaluating finite-sample performance become more subtle in dependent settings, further complicating the use of asymptotic approximations as a guide to inference. Given the prevalence of dependent data in economic applications, further development of practical diagnostics and implementation guidance tailored to this setting would be a welcome addition to the literature.

%Existing theory shows that cross-fitting and using estimated nuisance functions introduces no asymptotic loss relative to the oracle case where nuisance functions are known and the full sample is used to estimate the target parameter. These theoretical results provide implementable inference and perform well in many settings. However, a deeper understanding of the finite-sample behavior of DML---including tradeoffs in learner complexity, fold count, and cross-fitting repetition---would be valuable not just for refining our knowledge but in guiding practical implementation choices. 

\paragraph{Functional Restrictions.} A recent proposal for improving performance involves empirical calibration of nuisance function estimates, as developed by \citet{laanDoublyRobustInference2025}. Their approach adapts ideas from predictive modeling, requiring that estimated nuisance functions satisfy known properties. For example, if the nuisance function is $\eta_0(X) = \Ep[Y|X]$, we should not be able to improve mean squared prediction error by transforming $\eta_0(X)$. \citet{laanDoublyRobustInference2025} propose methods to enforce such calibration and show that, for a broad class of target parameters, DML estimators remain asymptotically normal even if only one nuisance estimator satisfies the usual convergence rate. More generally, using known functional restrictions can reduce the complexity of nuisance estimation, potentially improving performance even in complex settings. More work on incorporating economically motivated constraints seems promising, both as a practical aid and as a direction for methodological research.

\paragraph{Conclusion.}
In sum, while recognizing that important challenges remain, we believe DML is a valuable addition to the empirical researcher's toolkit. It offers a framework for incorporating machine learning methods into empirical analysis, which is increasingly important as economic data grow in richness and complexity. At the same time, part of our goal in this review is to underscore that DML is not a foolproof, mechanical procedure. Empirical results can depend sensitively on implementation choices, and careful diagnostics, robustness checks, and transparent reporting are essential. Viewed in this light, DML should be understood as a structured approach that, when used thoughtfully, can improve empirical practice, either as a mechanism for obtaining primary conclusions from complex data or as a systematic robustness check alongside more traditional methods.%At a minimum, DML provides a practical starting point for researchers presented with complex data and serves as a useful robustness check for results obtained from more traditional methods. As economic data become more complex and machine learning tools continue to advance, DML is likely to be increasingly important in applied research.

%%%%%%%%%%%%%%%%%%%%%%%%%%%%%%%%%%%%%%%%%%%%%%%%%%%%%%%%%%%%%%%%%%%%%%%%%%%%%%%%%%%%%%%%%
%%%%%%%%%%%%%%%%%%%%%%%%%%%%%%%%%%%%%%%%%%%%%%%%%%%%%%%%%%%%%%%%%%%%%%%%%%%%%%%%%%%%%%%%%
\clearpage

\appendix

\section*{Appendix}

We collect more technical discussions related to orthogonal scores in this appendix. In Appendix~\ref{app: neyman verification}, we illustrate verification of Neyman orthogonality, or the lack thereof, for the scores in Examples 1, 3, and 4. We present a general ``partialling out'' approach for constructing Neyman orthogonal scores from a given generic score function, which may not itself be Neyman orthogonal, in Appendix~\ref{app:no_construction}. Finally, we present additional examples of common target parameters and their Neyman orthogonal scores in Appendix~\ref{app:add_params}.

\section{Verification of Neyman Orthogonality}\label{app: neyman verification}

In this appendix, we verify Neyman orthogonality of the linear regression, linear IV, and AIPW scores presented in Examples 1, 3, and 4. We also verify that the IPW score from Example 4 does not satisfy Neyman orthogonality.

\bigskip

\noindent\textbf{Example 1 (continued). Neyman Orthogonality of the Linear Regression Score.}
In the linear regression score \eqref{eq: lm score}, the nuisance parameters $\eta_Y$ and $\eta_D$ are vectors. The second condition in \eqref{eq:neyman_def} then reduces to 
\begin{align*}
    \frac{\partial}{\partial \eta_Y} \Ep[m_{LM}(W;\theta_0,\eta)]\big\vert_{\eta = \eta_0} &= -\Ep[X(D-X'\eta_{D,0})] = 0_p \\
        \frac{\partial}{\partial \eta_D} \Ep[m_{LM}(W;\theta_0,\eta)]\big\vert_{\eta = \eta_0} &= 2\theta_0\Ep[X(D-X'\eta_{D,0})]-\Ep[X(Y-X'\eta_{Y,0})] = 0_p
\end{align*}
by applying \eqref{eq: lm normal}. $ $\hfill $\qed$

\bigskip

\noindent\textbf{Example 3 (continued). Neyman Orthogonality of the Linear IV Score.}
For the linear IV score \eqref{eq: iv}, the nuisance function is $\eta(Z)$ with true value $\eta_0(Z) = \Ep[D|Z]$. Considering a small perturbation, $\Delta\eta(Z) = \eta(Z) - \eta_0(Z)$, around $\eta_0(Z)$ yields
\begin{align*}
    \frac{\partial}{\partial \lambda}\Ep[m_{IV}(W;\theta_0,\eta_0(Z) + \lambda\Delta \eta(Z))]\big\vert_{\lambda = 0} = \Ep\left[(Y-\theta_0 D)\Delta\eta(Z)\right] = 0,
\end{align*}
where the last equality follows from $\Ep[\varepsilon|Z] = 0.$\footnote{Linear IV under conditional mean independence of $Z$ from $\varepsilon$ satisfies a much stronger condition than Neyman orthogonality in that the score equation at $\theta_0$ is globally insensitive to the nuisance function.} $ $\hfill $\qed$

\bigskip

\noindent\textbf{Example 4 (continued). Neyman Orthogonal Scores for the ATE.}
We now verify that the IPW score is not Neyman orthogonal, while the AIPW score is.

For the IPW score \eqref{eq: IPW}, the nuisance function is $\alpha(D,X)$ with true value $\alpha_0(D,X) = \frac{D}{r_0(X)} - \frac{1-D}{1-r_0(X)}$. Let $\Delta \alpha(D,X) = \alpha(D,X) - \alpha_0(D,X)$, then 
\begin{align*}
        \frac{\partial }{\partial \lambda} \Ep \left[m_{IPW}(W_i; \theta_0, \right. & \left. \alpha_0(D,X) + \lambda\Delta \alpha(D,X))\right]\big\vert_{\lambda=0} %\\
        %&= \Ep [Y  \Delta \alpha(D,X)] 
     = \Ep [\ell_0(D,X)
     \Delta \alpha(D,X)] \neq 0,
\end{align*} 
where $\ell_0(D,X) = \Ep[Y|D,X]$. Thus, the IPW score is not Neyman orthogonal.

The AIPW score \eqref{eq: AIPW} uses nuisance functions $\eta(D,X) = (\alpha(D,X),\ell(D,X))$ with true values $(\alpha_0(D,X),\ell_0(D,X))$ defined above. Let $\Delta \eta(D,X) = \eta(D,X) - \eta_0(D,X)$, then
\begin{align*}
\Ep[m_{AIPW}(W_i; &\theta_0, \eta_0(D,X) + \lambda\Delta\eta(D,X))] \\ &= -\lambda^2\Ep[\Delta \alpha(D, X) \Delta \ell(D, X)] \\
&\qquad + \Ep[(\alpha_0(D, X) + \lambda\Delta \alpha(D, X))(Y - \ell_0(D, X))] \\
&\qquad + \lambda\Ep[(\Delta \ell(1, X) - \Delta \ell(0, X) - \alpha_0(D, X) \Delta \ell(D, X))] \\
& \qquad + \Ep[\ell_0(1, X) - \ell_0(0, X)] - \theta_0 \\
& = -\lambda^2\Ep \left[\Delta \alpha(D, X) \Delta \ell(D,X) \right], 
\end{align*}
where the last equality 
follows from $\Ep[\ell_0(1, X) - \ell_0(0, X)] - \theta_0 = 0$ by the definition of the ATE, the definition of $\eta_0(D,X)$, and application of the law of expectations. Neyman orthogonality is then immediate:
$$
 \frac{\partial }{\partial \lambda}  \Ep\left[
m(W; \theta_0, \eta_0 + \lambda\Delta\eta)\right] \big\vert_{\lambda=0} = \frac{\partial }{\partial \lambda} \lambda^2  \big\vert_{\lambda=0} \Ep \left[\Delta \alpha(D, X) \Delta \ell(D,X) \right]= 0.
$$ 
$ $\hfill $\qed$

\section{Constructing Neyman Orthogonal Scores}\label{app:no_construction}

This appendix illustrates a ``partialling out'' approach to constructing Neyman orthogonal scores, generalizing the familiar approach of multiple linear regression discussed in Example~1. Throughout, we use $m$ to denote a generic score function and introduce the notation $\psi$ to denote a Neyman orthogonal score. This notational distinction helps clarify the construction of orthogonal scores from baseline moment conditions. We focus on scalar-valued target parameters $\theta_0$ for ease of exposition.

In many settings, we can readily obtain a moment condition $m$ that identifies $\theta_0$ as in \eqref{eq: score_identification}, i.e., $\Ep[m(W;\theta_0,\eta_0)] = 0$, but is not Neyman orthogonal. Often, the nuisance parameter in such cases is a vector of conditional expectation functions. That is, $\eta_0 = (\gamma_0^{(h)})_{h=1}^H$ where, for each $h$, $\gamma_0^{(h)}(B^{(h)})= \Ep[A^{(h)}\vert B^{(h)}]$ for some subvectors $A^{(h)}$ and $B^{(h)}$ of $W$. Further, the score function $m$ typically depends on $\eta_0$ only through its value $\eta_0(W)$.

Consider, for example, the ATE defined in \eqref{eq:def_ATE} 
with corresponding IPW score \begin{align*}
    m_{IPW}(W; \theta, \eta)=\frac{D Y}{\gamma^{(1)}(X)} -\frac{(1-D)Y}{1-\gamma^{(1)}(X)} -\theta,
\end{align*}
where the nuisance parameter $\eta= \gamma^{(1)}$ takes its true value at $\gamma_{0}^{(1)}(X)= \Ep[D\vert X]$. In terms of the general notation, this corresponds to $A^{(1)}= D$ and  $B^{(1)}= X$. Clearly, the IPW score depends on the nuisance parameter only through its value $\gamma_{0}^{(1)}(X)$.

\citet{newey94} shows how to calculate the impact of nuisance estimation for this kind of moment condition and highlights that these calculations facilitate the construction of Neyman orthogonal scores by projecting this impact onto covariates---a generalization of the partialling out approach in multiple linear regression discussed in Example 1. See also \citet{chernozhukov2018}, \citet{chernozhukov2021automatic},  \citet{chernozhukov2022locally}, and \citet{kennedy:review} for further discussion and approaches to construct orthogonal scores. 

Heuristically, the first-order impact of bias in the $h^{\textnormal{th}}$ nuisance parameter $\gamma^{(h)}$ is 
\begin{align}
\begin{split}\label{eq:neyman_def_chain}
\frac{\partial}{\partial \lambda} m(W; \theta_0, \gamma_{0}^{(1)}, \ldots, \gamma_{0}^{(h)} + \lambda\Delta \gamma^{(h)}, \ldots, \gamma_{0}^{(H)})\big\vert_{\lambda=0} = m_{h}(W; \theta_0, \eta_0) \Delta \gamma^{(h)}(B^{(h)})
\end{split}
\end{align}
where the equality follows from the chain rule, and we use $\Delta \gamma^{(h)} = \gamma_{0}^{(h)} - \gamma^{(h)}$ to denote a deviation of the nuisance from its true value and $m_h$ to denote the partial derivative of $m$ with respect to the value of $\gamma^{(h)}$. 

To correct for the impact of the bias in the $h^{\textnormal{th}}$ nuisance parameter, define the adjustment factor $\alpha_{0}^{(h)}(B^{(h)})$ as the projection of $m_h(W; \theta_0, \gamma_0)$ onto covariates $B^{(h)}$:
\begin{align}\label{eq:alpha_proj}
    \alpha_{0}^{(h)}(B^{(h)})= \Ep[m_h(W; \theta_0, \eta_0)\vert B^{(h)}].
\end{align}
By the law of iterated expectations, $\alpha_{0}^{(h)}$ satisfies the orthogonality condition \begin{align}\label{eq:LIE_projectionalpha}
    \Ep[\alpha_{0}^{(h)}(B^{(h)})(A^{(h)} - \gamma_{0}^{(h)}(B^{(h)}))] = 0.
\end{align}
Note that the construction of adjustment factors closely parallels that of the best linear predictors in Example~1, which satisfy a weaker---but otherwise analogous---orthogonality condition \eqref{eq: lm normal}. The key difference is that \eqref{eq:alpha_proj} relaxes the restriction of best \emph{linear} predictors, considering instead the broader class of conditional expectation functions.

A Neyman orthogonal score can then be constructed as \begin{align}\label{eq:psi_correction}
    \psi(W; \theta, \eta) = m\left(W; \theta, (\gamma^{(h)})_{h=1}^H\right) + \sum_{h=1}^H \alpha^{(h)}(B^{(h)})(A^{(h)} - \gamma^{(h)}(B^{(h)})),
\end{align}
where $\eta=(\gamma^{(h)}, \alpha^{(h)})_{h=1}^H$ is the combined nuisance parameter with true value $\eta_0=(\gamma_0^{(h)}, \alpha_0^{(h)})_{h=1}^H$. To verify that the score in \eqref{eq:psi_correction} indeed satisfies Neyman orthogonality, note first that the orthogonality condition \eqref{eq:LIE_projectionalpha} implies that $\psi$ and $m$ identify the same target parameter. Then, by the chain rule 
\begin{align*}
    \frac{\partial}{\partial \lambda}\big\{\Ep\left[\psi(W; \right.  \left. \theta_0, \eta_0 + \lambda\Delta\eta)\right]\big\}\big\vert_{\lambda=0} 
    =&\sum_{h=1}^H \Ep\left[\left(m_{h}(W; \theta_0, \gamma_0)  - \alpha_{0}^{(h)}(B^{(h)})\right)\Delta \gamma^{(h)}(B^{(h)})\right] \\
    &\quad +\sum_{h=1}^H \Ep\left[\Delta \alpha^{(h)}(B^{(h)}) \left(A^{(h)}-\gamma_0^{(h)}(B^{(h)})\right)\right]
    =0,
\end{align*}
where the final equality follows from \eqref{eq:alpha_proj} and the law of iterated expectations. Mirroring the familiar approach in multiple linear regression (Example~1) or partially linear regression (Example~2), the score \eqref{eq:psi_correction} is thus constructed to ``partial out'' the first-order impact of estimating the nuisance functions: for each $h$, small errors in either $\gamma_0^{(h)}(B^{(h)})$ or $\alpha_{0}^{(h)}(B^{(h)})$ can be offset by the other, making estimation more robust.

Returning to the example of the average treatment effect and its IPW score, we can thus compute the impact of propensity score estimation as 
\begin{align*}
    \frac{\partial}{\partial \lambda} m(W; \theta_0, \gamma_{0}^{(1)} + \lambda\Delta \gamma^{(1)})\big\vert_{\lambda=0} %&= -\frac{D_iY_i\Delta \gamma^{(1)}(X_i) }{\gamma_{0}^{(1)}(X_i)^2} - \frac{(1-D_i)Y_i\Delta \gamma^{(1)}(X_i)}{(1-\gamma_{0}^{(1)}(X_i))^2} 
    = m_1(W; \theta_0, \gamma_{0}^{(1)})\Delta \gamma^{(1)}(X)%\\
    %&=\left(-\frac{D_iY_i}{\gamma_{0}^{(1)}(X_i)^2} - \frac{(1-D_i)Y_i}{(1-\gamma_{0}^{(1)}(X_i))^2}\right)\Delta \gamma^{(1)}(X_i),
\end{align*}
where %\begin{align*}
    $m_1(W; \theta_0, \gamma_{0}^{(1)}) = -\frac{D Y}{\gamma_{0}^{(1)}(X)^2} - \frac{(1-D)Y}{(1-\gamma_{0}^{(1)}(X))^2}.$
%\end{align*}
We then construct the adjustment factor as \begin{align*}
    \alpha_{0}^{(1)}(X) &= \Ep[m_1(W; \theta_0, \gamma_{0}^{(1)})\vert X] = -\frac{\Ep[Y\vert D=1, X]}{\gamma_{0}^{(1)}(X)} - \frac{\Ep[Y\vert D=0, X]}{1-\gamma_{0}^{(1)}(X)},
\end{align*}
where the final equality follows from the law of total probability. Combining provides a Neyman orthogonal score for the ATE:
\begin{align*}
   \psi(W; \theta, \eta) &= \frac{D Y}{\gamma^{(1)}(X)} -\frac{(1-D)Y}{1-\gamma^{(1)}(X)} -\theta  +\alpha^{(1)}(X)(D - \gamma^{(1)}(X)).
\end{align*}
%This score function differs from the alternative AIPW score for the ATE in Example~4 in its parameterization of the nuisance parameter. 
After replacing $\alpha^{(1)}$ with its representation in terms of conditional expectations and some algebra, this expression reduces to the AIPW score for the ATE in Example~4.

\begin{remark}[Adjustment Factors $\alpha^{(h)}_0$ are Riesz Representers]\label{remark:RRs}
The orthogonal score construction described above is closely connected to the concept of the \emph{Riesz representer}, a fundamental object in functional analysis. In particular, the adjustment factors \eqref{eq:alpha_proj} are Riesz representers for the expected first-order impact of bias in the $h^{\textnormal{th}}$ nuisance function. To see this, note that taking expectations of the first-order bias \eqref{eq:neyman_def_chain} defines a functional $\phi^{(h)}:\Gamma^{(h)} \to \mathbbm{R}$ for each $h$,\begin{align}\label{eq:functional_expectedbias}
    \phi^{(h)}(f) \equiv \Ep\left[ m_{h}(W; \theta_0, \eta_0) f(B^{(h)})\right],
\end{align}
where $\Gamma^{(h)}$ denotes the appropriate functional space of $\gamma_0^{(h)}$ and $f\in\Gamma^{(h)}$ denotes any potential nuisance bias function $\Delta \gamma^{(h)}$.
Under appropriate conditions, the Riesz representation theorem implies existence of a unique $\alpha^{(h)}_0\in\Gamma^{(h)}$ such that for any $f\in \Gamma^{(h)}$, $\phi^{(h)}(f)=\Ep[\alpha^{(h)}_0(W)f(B^{(h)})]$. The law of iterated expectations then implies that these so-called Riesz representers are given by the adjustment factors \eqref{eq:alpha_proj}.

The insight that the adjustment factors are Riesz representers facilitates construction of Neyman orthogonal scores, even when an explicit expression for $\alpha_0$ is unavailable. This is key to recent DML approaches that rely on implicitly defined Riesz representers \citep[e.g.,][]{chernozhukov2021automatic, chernozhukov2022automaticECMA, CNS:globalandlocal, hirschberg:wager}.%, with the last reference building on the former.
\end{remark}

\begin{remark}[Calibrated Riesz Representers]
Viewing the adjustment factors $\alpha_0^{(h)}$ as Riesz representers as highlighted in Remark~\ref{remark:RRs} allows researchers to leverage properties implied by the Riesz representation theorem that can further improve finite sample behavior of DML estimators. As an example, note that taking $f=\alpha^{(h)}_0$ in \eqref{eq:functional_expectedbias} where $\alpha^{(h)}_0\in\Gamma^{(h)}$, we have $
    \Ep\left[ m_{h}(W; \theta_0, \gamma_0) \alpha^{(h)}_0(B^{(h)})\right] = \Ep[\alpha^{(h)}_0(B^{(h)})^2].$
This moment condition holds in the population, but its sample analog based on the DML estimator in Algorithm~\ref{alg: DML} generally does not. To ensure that the sample analog holds as well, we can calibrate an initial estimate of the adjustment factor $\hat{\alpha}^{(h)}$ by introducing a scalar $\hat{\iota}_h$ and defining $\hat{\alpha}^{(h)}_{\text{cal}} = \hat{\iota}_h \hat{\alpha}^{(h)}$. Assuming $\hat{\iota}_h \neq 0$, $\hat{\iota}_h$ is set to
\begin{align*}
    \En\left[ \hat{m}_{h}(W) (\hat{\iota}_k \hat{\alpha}^{(h)}(B^{(h)})) \right] = \En\left[ (\hat{\iota}_h \hat{\alpha}^{(h)}(B^{(h)}))^2 \right] \: \Leftrightarrow \:  \hat{\iota}_h = \frac{\En[\hat{m}_{h}(B^{(h)}) \hat{\alpha}^{(h)}(B^{(h)})]}{\En[\hat{\alpha}^{(h)}(B^{(h)})^2]}.
\end{align*}
%Calibration did not qualitatively affect the conclusions presented in the empirical analyzes presented in this paper.
See \citet{van2024doubly} for further discussion of \textit{calibrated} DML.
\end{remark}

\section{Neyman Orthogonal Scores for Additional Common Target Parameters}\label{app:add_params}

We now provide additional examples of common target parameters and their Neyman orthogonal scores. For ease of exposition, we categorize these parameters as treatment effect parameters (Section~\ref{app:TE_params}), regression parameters (Section~\ref{app:reg_params}), and fixed effect regression parameters (Section~\ref{app:fe_reg_params}). We use the same notation and structure as in Appendix \ref{app:no_construction}.

\subsection{Treatment Effect Parameters}\label{app:TE_params}

Consider $W = (Y, D, Z, X)$ where $Y$ is a scalar outcome, $D$ is a discrete treatment, $Z$ is a binary instrument, and $X$ is a vector of controls. Under standard assumptions \citep[e.g.,][]{ImbensRubin2015}, the following objects are well-defined causal quantities.

\subsubsection{Weighted Average Potential Outcome}

The weighted average potential outcome corresponding to treatment level $d$ is 
\begin{align*}
   \theta_0 = \Ep[\omega(X)\Ep[Y\vert D = d, X]],
\end{align*}
for some known weighting function $\omega$. The corresponding IPW score is \begin{align*}
  m(W; \theta, \gamma^{(1)}) = \frac{\mathbbm{1}\{D=d\}Y}{\gamma^{(1)}(X)}\omega(X)-\theta,
\end{align*}
where the nuisance parameter $\gamma^{(1)}$ has true value $\gamma_{0}^{(1)}(X) = \Ep[\mathbbm{1}\{D=d\}\vert X]$. The corresponding correction term is given by \begin{align*}
   \alpha^{(1)}_{0}(X) = -\frac{\Ep[Y\vert D=d, X]\omega(X)}{\Ep[\mathbbm{1}\{D=d\}\vert X]}.
\end{align*}

\subsubsection{Average Treatment Effect on the Treated}

The average treatment effect on the treated for a binary treatment $D$ is 
\begin{align*}
   \theta_0 = \Ep\left[\Ep[Y \vert D = 1, X] - \Ep[Y \vert D = 0, X]\vert D = 1\right],
\end{align*}
with corresponding IPW score \begin{align*}
   m(W; \theta, \gamma^{(1)}, \gamma^{(2)}) = \frac{D Y}{\gamma^{(2)}} - \frac{\gamma^{(1)}(X)(1-D)Y}{\gamma^{(2)}(1-\gamma^{(1)}(X))} - \frac{D}{\gamma^{(2)}}\theta,
\end{align*}
where the nuisance parameters $\gamma^{(1)}$ and $\gamma^{(2)}$ take true values at $\gamma_{0}^{(1)}(X)= \Ep[D\vert X]$ and $\gamma_{0}^{(2)} = \Ep[D]$. The corresponding correction terms are given by 
\begin{align*}
    \alpha_{0}^{(1)}(X)= -\frac{1}{\Ep[D]}\left(\frac{\Ep[D\vert X]\Ep[Y\vert D=0, X]}{1 - \Ep[D\vert X]} + \Ep[Y\vert D=0, X]\right),\quad
    \alpha_{0}^{(2)} = 0.
\end{align*}

\subsubsection{Local Average Treatment Effect}

The local average treatment effect (LATE) for a binary instrument $Z$ is defined as 
\begin{align*}
   \theta_0 = \frac{\Ep\left[\Ep[Y \vert Z = 1, X] - \Ep[Y \vert Z = 0, X]\right]}{\Ep\left[\Ep[D \vert Z = 1, X] - \Ep[D \vert Z = 0, X]\right]} ,
\end{align*}
with corresponding IPW score 
\begin{align*}
  m(W; \theta, \gamma^{(1)}) = \frac{Z Y}{\gamma^{(1)}(X)} -\frac{(1-Z)Y}{1-\gamma^{(1)}(X)} - \theta\left(\frac{Z D}{\gamma^{(1)}(X)} -\frac{(1-Z)D}{1-\gamma^{(1)}(X)}\right),
\end{align*}
where the nuisance parameter $\gamma^{(1)}$ takes true value at $\gamma^{(1)}_{0}(X)= \Ep[Z\vert X]$. The corresponding correction term is given by
\begin{align*}
    \alpha^{(1)}_{0}(X)= -\frac{\Ep[Y\vert Z=1, X]}{\Ep[Z\vert X]} - \frac{\Ep[Y\vert Z=0, X]}{1-\Ep[Z\vert X]} + \theta_0\left(\frac{\Ep[D\vert Z=1, X]}{\Ep[Z\vert X]} + \frac{\Ep[D\vert Z=0, X]}{1-\Ep[Z\vert X]}\right).
\end{align*} 

\subsection{Regression Parameters}\label{app:reg_params}
Consider $W = (Y, D, Z, X)$ where $Y$ is a scalar-valued outcome, $D$ is a vector of variables of interest, $Z$ is a vector of instruments, and $X$ is a vector of controls.

\subsubsection{Partially Linear Regression and Partially Linear IV}\label{app:plm}

Consider the instrumental variable (IV) regression
\begin{align*}
    Y = D^\top \theta_0 + g_0(X) + \varepsilon, 
\end{align*}
where the target parameter $\theta_0$ and confounding function $g_0(\cdot)$ are defined through the orthogonality restrictions $\Ep[Z\varepsilon] = 0$ and $\Ep[\varepsilon \vert X] =0$, and the IV relevance condition that $\Ep[\Cov(Z, D\vert X)]
$ has full column rank. This setting corresponds to a scenario where a researcher has a known set of instruments, $Z$, that are taken to satisfy the exclusion restriction only after conditioning on controls $X$ and does not wish to impose the functional form in which confounds enter the model.
Further note that we recover partially linear regression by setting $Z = D$. Solving for $g_0$ and substituting, we obtain the vector-valued score \begin{align*}
     m(W; \theta, \gamma^{(1)}, \gamma^{(2)}) = Z(Y - \gamma^{(1)}(X) -\theta^\top(D -\gamma^{(2)}(X))),
\end{align*}
where the nuisance parameters take true value at $\gamma^{(1)}_{0}(X) = \Ep[Y\vert X]$ and $\gamma^{(2)}_{0}(X) = \Ep[D\vert X]$. The corresponding correction terms are given by 
\begin{align*}
    \alpha^{(1)}_{0}(X) = -\Ep[Z\vert X], \quad 
    \alpha^{(2)}_{0}(X) = \Ep[Z\vert X]\theta^\top.
\end{align*}

\subsubsection{Flexible Partially Linear Instrumental Variables}
Consider the IV regression
\begin{align*}
    Y = D^\top \theta_0 + g_0(X) + \varepsilon, 
\end{align*}
where the target parameter $\theta_0$ and the confounding function $g_0(\cdot)$ are defined through the orthogonality restrictions $\Ep[\varepsilon \vert Z, X] = 0$, and the IV relevance condition that $\Ep[\Var(\Ep[D\vert Z, X]\vert X)]
$ is positive definite. This setting differs from that considered in Section \ref{app:plm} in that the orthogonality condition $\Ep[\varepsilon \vert Z, X] = 0$ is stronger. It implies that any function of $(Z, X)$ can be used as a valid instrument. We consider the optimal instrument under homoskedasticity ($\Ep[D \vert Z, X]$).%\footnote{Other targets could also be considered.}

Using the fact that $\Ep[\varepsilon \vert Z, X] = 0$ implies $\Ep[\varepsilon \vert X] = 0$ by the law of iterated expectations, we can solve for $g_0$. Further, note that $\Ep[\varepsilon \vert Z, X] = 0$ implies $\Ep[\gamma^{(3)}_{0}(Z, X)\varepsilon] = 0$ for $\gamma^{(3)}_{0}(Z, X) = \Ep[D \vert Z, X]$. This motivates the vector-valued score \begin{align*}
   m(W; \theta, \gamma^{(1)}, \gamma^{(2)}, \gamma^{(3)}) = \gamma^{(3)}(Z, X)(Y - \gamma^{(1)}(X) -\theta^\top(D -\gamma^{(2)}(X))),
\end{align*}
where the nuisance parameters $\gamma^{(1)}$, $\gamma^{(2)}$, and $\gamma^{(3)}$ take true values, respectively, at $\gamma^{(1)}_{0}(X) = \Ep[Y\vert X]$, $\gamma^{(2)}_{0}(X) = \Ep[D\vert X]$, and $\gamma^{(3)}_{0}(Z, X) = \Ep[D\vert Z, X]$.
The corresponding correction terms are given by \begin{align*}
    \alpha^{(1)}_{0}(X) = -\Ep[D\vert X], \quad 
    \alpha^{(2)}_{0}(X) = \Ep[D\vert X]\theta^\top,\quad 
    \alpha^{(3)}_{0}(Z, X) = 0.
\end{align*}

\subsection{Fixed Effect Regression Parameters}\label{app:fe_reg_params}
Consider $W_i = (Y_{i,t}, D_{i,t}, Z_{i,t}, X_{i,t})^T_{t=0}$ where $t$ denotes a secondary dimension (e.g., time), $Y_{i,t}$ is a scalar-valued outcome, $D_{i,t}$ is a vector of variables of interest, $Z_{i,t}$ is a vector of instruments, and $X_{i,t}$ is a vector of controls. We explicitly index by $i$ and $t$ to introduce cross-sectional heterogeneity via individual fixed effects. It is convenient to define the first difference operator $\Delta A_{i,t} = A_{i,t} - A_{i,t-1}$ for random variables $A_{i,t}$ and $A_{i,t-1}$.

Consider the IV regression with fixed effects $\iota_i$
\begin{align*}
    Y_{i,t} = D_{i,t} ^\top \theta_0 + g^{(t)}_{0}(X_{i,t}) + \iota_i + \varepsilon_{i,t}, \quad \forall t =0, 1, \ldots, T,
\end{align*}
where the target parameter $\theta_0$ and the differenced confounding functions $\{\Delta g^{(t)}_{0}\}_{t=0}^T$ are defined through the orthogonality restrictions $\Ep[\sum_{t=1}^T \Delta Z_{i,t}\Delta\varepsilon_{i,t}] = 0$, $\Ep[\Delta\varepsilon_{i,t} \vert X_{i,t}, X_{i,t-1}] =0, \forall t \in \{1, \ldots, T\}$, and the IV relevance condition that $\Ep[\Cov(\Delta Z_{i,t}, \Delta D_{i,t}\vert X_{i,t}, X_{i,t-1})]$ has full column rank for at least some $t \in \{1, \ldots, T\}$. Note that we recover fixed effects partially linear regression by setting $(Z_{i,t})^T_{t=1} = (D_{i,t})^T_{t=1}$.
Solving for $\Delta g^{(t)}_{0}$ and substituting, we obtain the vector-valued score \begin{align*}
     m(W_i; \theta, \{\gamma^{(1t)}, \gamma^{(2t)}\}_{t=1}^T) = \sum_{t=1}^T \Delta Z_{i,t}(\Delta Y_{i,t} - \gamma^{(1t)}(X_{i,t}, X_{i,t-1}) -\theta^\top(\Delta D_{i,t} -\gamma^{(2t)}(X_{i,t}, X_{i,t-1}))),
\end{align*}
where the nuisance parameters $\{\gamma^{(1t)}, \gamma^{(2t)}\}_{t=1}^T$ take true value at $\gamma^{(1t)}_0(X_{i,t}, X_{i,t-1}) = \Ep[\Delta Y_{i,t}\vert X_{i,t}, X_{i,t-1}]$ and $\gamma^{(2t)}_0(X_i) = \Ep[\Delta D_{i,t}\vert X_{i,t}, X_{i,t-1}]$, for all $t \in \{1, \ldots, T\}$. The corresponding correction terms are given by \begin{align*}
    \alpha^{(1t)}_{0}(X_{i,t}, X_{i,t-1}) = -\Ep[\Delta Z_{i,t}\vert X_{i,t}, X_{i,t-1}], \quad 
    \alpha^{(2t)}_{0}(X_{i,t}, X_{i,t-1}) = \Ep[\Delta Z_{i,t}\vert X_{i,t}, X_{i,t-1}]\theta^\top.
\end{align*}

%%%%%%%%%%%%%%%%%%%%%%%%%%%%%%%%%%%%%%%%%%%%%%%%%%%%%%%%%%%%%%%%%%%%%%%%%%%%%%%%%%%%%%%%%
%%%%%%%%%%%%%%%%%%%%%%%%%%%%%%%%%%%%%%%%%%%%%%%%%%%%%%%%%%%%%%%%%%%%%%%%%%%%%%%%%%%%%%%%%
\printbibliography

\end{document}

%% file: notation.tex
\newcommand{\iid}{\emph{i.i.d.}\ }

\newcommand{\be}{\begin{eqnarray}}
	\newcommand{\ee}{\end{eqnarray}}

\newcommand{\ba}{\begin{array}}
	\newcommand{\ea}{\end{array}}
\newcommand{\bs}{\begin{align}\begin{split}\nonumber}
		\newcommand{\bsnumber}{\begin{align}\begin{split}}
				\newcommand{\es}{\end{split}\end{align}}

		\renewcommand{\[}{\left[}
		
		\renewcommand{\hat}{\widehat}

		\newcommand{\En}{{\mathbb{E}_n}}
		\newcommand{\Ep}{\mathrm{E}}%{\mathbf{E}}
		\newcommand{\Var}{\text{Var}}
        \newcommand{\Cov}{\text{Cov}}

		\def\RR{{\rm I\kern-0.18em R}}

		\renewcommand{\Pr}{{\mathrm{P}}}

%% file: theory_envs.tex
\usepackage{amsthm}
\usepackage{thmtools}
\usepackage[framemethod=TikZ]{mdframed}

\newtheoremstyle{plainbox}
  {}   % Space above
  {}   % Space below
  {\normalfont}    % Body font (normalfont instead of italic)
  {}          % Indent amount
  {\bfseries} % Theorem head font
  {}         % Punctuation after theorem head
  {\newline}      % Space after theorem head
  {}          % Theorem head spec

\theoremstyle{plainbox}

\mdfdefinestyle{theorembox}{
    linewidth=1pt,
    innertopmargin=1pt,
    innerbottommargin=\topskip,
    innerrightmargin=10pt,
    innerleftmargin=10pt,
}

\mdfdefinestyle{nobox}{
    linewidth=0pt,
    innertopmargin=0pt,
    innerbottommargin=0pt, %\topskip,
    innerrightmargin=0pt,
    innerleftmargin=0pt,
}

\newmdtheoremenv[style=theorembox]{definition}{Definition}
\newmdtheoremenv[style=theorembox]{theorem}{Theorem}
\newmdtheoremenv[style=theorembox]{lemma}{Lemma}
\newmdtheoremenv[style=theorembox]{remark}{Remark}
\newmdtheoremenv[style=nobox]{example}{Example}

\newmdtheoremenv[style=theorembox]{corollary}{Corollary}
\newmdtheoremenv[style=theorembox]{proposition}{Proposition}
\newmdtheoremenv[style=theorembox]{assumption}{Assumption}
\newmdtheoremenv[style=theorembox]{recommendation}{Recommendation}

\newmdtheoremenv[style=nobox,continued]{examplecont}{Example}

\AtBeginEnvironment{example}{\pushQED{\qed}}
\AtEndEnvironment{example}{\popQED}

%% file: tables/hrs/attgt_ranger2.tex
\begin{tabular}{cccccccccc}
\toprule \midrule
& & \multicolumn{5}{c}{\it Cross-fitting Repetitions} & & & \\
\cmidrule(lr){3-7}
\makecell{Wave first \\ hospitalized} & Wave & Rep.\ 1 & Rep.\ 2 & Rep.\ 3 & Rep.\ 4 & Rep.\ 5 & \makecell{Median \\ aggregate} & Const.\ & AIPW \\
& & (1) & (2) & (3) & (4) & (5) & (6) & (7) & (8)\\
\midrule
\multirow{8}[0]{*}{8} & \multirow{2}[0]{*}{7} & 0 & 0 & 0 & 0 & 0 & 0 & 0 & 0 \\
      &       & - & - & - & - & - & - & - & - \\
      & \multirow{2}[0]{*}{8} & 2467.6 & 2478 & 2442.9 & 2592.9 & 2564.7 & 2478 & 3028.6 & 2200.5 \\
      &       & (794.9) & (779.6) & (806.4) & (790.5) & (798.4) & (798.8) & (913.5) & (839) \\
      & \multirow{2}[0]{*}{9} & 626.8 & 466 & 559.2 & 483 & 519.9 & 519.9 & 1247.7 & -14.3 \\
      &       & (591.9) & (628.3) & (601.6) & (642.8) & (591.9) & (602.8) & (860.7) & (681) \\
      & \multirow{2}[0]{*}{10} & 723.7 & 803 & 766.6 & 614.8 & 691.8 & 723.7 & 800.1 & 998.9 \\
      &       & (582.6) & (580.9) & (598.3) & (599.3) & (616.3) & (599.9) & (1007.5) & (570.2) \\
& & & & & & & & & \\
\multirow{8}[0]{*}{9} & \multirow{2}[0]{*}{7} & 1405.6 & 1358.5 & 1412 & 1333.2 & 1324.4 & 1358.5 & 170 & 1717.8 \\
      &       & (1022.7) & (997.5) & (992.7) & (969.2) & (965.2) & (994.1) & (1128.4) & (1324.8) \\
      & \multirow{2}[0]{*}{8} & 0 & 0 & 0 & 0 & 0 & 0 & 0 & 0 \\
      &       & - & - & - & - & - & - & - & - \\
      & \multirow{2}[0]{*}{9} & 3485.6 & 3280.4 & 3218.3 & 3379.2 & 3204.5 & 3280.4 & 3324.4 & 3789.9 \\
      &       & (1000.4) & (941.9) & (931.1) & (953.2) & (923.3) & (941.9) & (958.8) & (1338.6) \\
      & \multirow{2}[0]{*}{10} & 983.1 & 936.9 & 1059.8 & 912.5 & 904.6 & 936.9 & 106.8 & 2533.6 \\
      &       & (431.3) & (427.9) & (432.3) & (436.5) & (425.5) & (433.7) & (650.7) & (421.7) \\
& & & & & & & & & \\
\multirow{8}[0]{*}{10} & \multirow{2}[0]{*}{7} & -1248 & -1783.6 & -1856.9 & -577.4 & -1491.7 & -1491.7 & 591 & 1221.5 \\
      &       & (2203.2) & (2473.6) & (2624.7) & (1661.1) & (2223.3) & (2223.3) & (1268.9) & (1126.7) \\
      & \multirow{2}[0]{*}{8} & 249.5 & 237.9 & 223.2 & 235.4 & 209 & 235.4 & 410.6 & 246.7 \\
      &       & (1016.2) & (1015.9) & (1019.2) & (1019.2) & (1022.7) & (1019.2) & (1027.1) & (1026.2) \\
      & \multirow{2}[0]{*}{9} & 0 & 0 & 0 & 0 & 0 & 0 & 0 & 0 \\
      &       & - & - & - & - & - & - & - & - \\
      & \multirow{2}[0]{*}{10} & 2731.1 & 2710.5 & 2690.2 & 2343.6 & 2480.2 & 2690.2 & 3091.5 & 3796.1 \\
      &       & (1215.2) & (1171.4) & (1260.2) & (1336.2) & (1260.5) & (1260.2) & (995.4) & (931.6) \\
\midrule
\bottomrule
\end{tabular}